\documentclass[pra,aps,showpacs,floatfix,twocolumn]{revtex4-1}
\pdfoutput=1
\usepackage{graphicx}
\usepackage{epstopdf}
\usepackage{wrapfig}
\usepackage{subfigure}
\usepackage{amsmath}
\usepackage{float}
\usepackage{amsmath}
\usepackage{microtype}
\usepackage{braket}
\usepackage{physics}
\usepackage{framed}
\usepackage[version=3]{mhchem}
\usepackage{hyperref} 
\usepackage{datetime}
\usepackage{ctable}
\usepackage[outline]{contour}

\newcommand{\me}{\mathrm{e}} 

\begin{document}
\title{Entanglement and its relationship to classical dynamics}
\author{Joshua B. Ruebeck, Jie Lin, and Arjendu K. Pattanayak}
\date{\today}
\affiliation{Department of Physics and Astronomy,
Carleton College, Northfield, Minnesota 55057, USA
}
\begin{abstract}
  We present an analysis of the entangling quantum kicked top
  focusing on the few qubit case and the initial condition dependence
  of the time-averaged entanglement $S_Q$ for spin-coherent states. We
  show a very strong connection between the classical phase space and
  the initial condition dependence of $S_Q$ even for the extreme case
  of two spin-$1/2$ qubits. This correlation is not related directly
  to chaos in the classical dynamics. We introduce a measure of
  the behavior of a classical trajectory which correlates far better
  with the entanglement and show that the maps of classical and
  quantum initial-condition dependence are both organized around the
  symmetry points of the Hamiltonian. We also show clear
  (quasi-)periodicity in entanglement as a function of number of kicks
  and of kick strength.
\end{abstract}

\maketitle 
\section{Introduction}
The relationship between the entanglement of a nonlinear quantum
system and the dynamics of its chaotic classical limit is deeply
intriguing since entanglement is quintessentially quantal and chaos
quintessentially classical. It has been extensively studied both
theoretically and experimentally for over two
decades~\cite{Furuya,Miller,arul,ghosesanders,wang,ghosesilberfarb,collin,jessen,matzkin}
and continues to be investigated today~\cite{pedram,swingle}.

One paradigmatic model system is the `kicked top,' consisting of a
nonlinearly evolving spin composed of $2j$ qubits and with total spin
$\vec{J}$.  The quantum behavior is usually mapped by studying the
entanglement dynamics of spin-coherent states initialized at various
locations in the phase space.  While different measures of the quantum
entanglement can be studied, the standard analysis considers the
time-dependent entanglement between any one qubit and the other $2j-1$
qubits.  This behavior is then compared with the classical point
dynamics of initial conditions corresponding to the locations of the
centroids of the spin-coherent states. Studies of this system have
considered systems with the quantum number ranging from $j=8$ up to
$j \approx 250$.

The common wisdom about the broad characteristics of the system
behavior can be summarized as follows: If a spin-coherent state has an
initial centroid location such that the corresponding classical
trajectory is chaotic, then (a) the quantum entanglement between the
subsystems depends on the classical largest Lyapunov exponent
$\lambda$ (which measures the degree of classical chaos), and moreover
follows changes in the behavior of $\lambda$ with system parameters,
and (b) more generally, the asymptotic entanglement and the
time-averaged entanglement for these `chaotic' initial states is
significantly greater than for states with initial centroids
corresponding to regular classical trajectories. Finally, it is
understood that (c) this `entanglement as quantum signature of
classical chaos' becomes more distinct as the number of spins $j$
increases, that is, as the effective $\hbar$ decreases in the
correspondence limit.  The reason for this connection is argued
broadly as follows: Classically chaotic initial conditions explore
phase-space more widely. Thus, if a quantum system corresponding to a
classically chaotic initial condition similarly explores Hilbert space
widely, and given that the generic Hilbert space state is entangled,
the average entanglement is consequently greater for such a quantum
system.

Dissenters from this consensus include Lombardi and
Matzkin~\cite{matzkin} who have argued with specific counterexamples
that high quantum entanglement can occur for initial conditions with
centroids initialized in classically regular regions. These authors
further compare the entanglement with an analagous quantity for a
classical probability distribution, deriving from the premise that the
classical and quantum (expectation value) dynamics agree with each
other for longer times for classical probability distributions than
for individual classical initial conditions. Unfortunately, the
classical distribution calculations are computationally expensive,
prohibiting a full scan of the phase space and a verification of this
idea. It is worth noting that all arguments to date have evoked the 
correspondence principle in explaining the relationship between the 
classical and quantum behavior. Further, statements about 
`correlations' between measures have not been quantified, and 
rely on the visual similiarity of various figures.

In this paper, we present results from a somewhat different
perspective on this issue.  We work at small $j$, corresponding to
recent experiments~\cite{pedram} and focusing in particular on two
coupled spin-$1/2$ qubits ($j=1$). For this system we are able to
analytically calculate the infinite-time-averaged quantum entanglement
$S_Q$ of initial spin coherent states. We see that $S_Q$ depends
strongly on the initial location of the spin coherent state.  We also
see that the geometry emerging from plots of the initial condition
dependence of $S_Q$ correlates strongly with the geometry of the
classical phase-space even for this extreme case where the quantum and
classical trajectories disagree immediately and the correspondence
principle cannot be evoked. However, high $S_Q$ does not correlate
with classical chaos. In particular, we see that classically regular
dynamics corresponds to either high or low entanglement, depending on
the properties of the orbit, while classically chaotic dynamics
correspond to entanglement levels about halfway between these
extremes.

In order to better explore and understand this unusual result, we
first systematize the so-far loose notion of the correlation between
the various functions of initial conditions that we use to
characterize the systems. That is, the strength of correlations
between various measures is quantified using a generalized
Kullback-Liebler distance rather than the usual visual
inspection. Secondly, we introduce a measure $I_C$ of the `ignorance'
associated with the time averaged location of a classical
trajectory. $I_C$ incorporates insights similar to Ref.~\cite{matzkin}
about orbit delocalization but is significantly easier to compute. We
see that $I_C$ correlates well with $S_Q$ across a wide range of
system dynamical behaviors, and certainly does better than any attempt
to correlate the entanglement with measures of chaos.  We argue that
the roots of this correlation may be traced to the fact that $S_Q$ is
equal to the sum of `diagonal' ($I_Q$) and `off-diagonal' ($R_Q$)
matrix elements of the angular momentum operators computed with the
Floquet eigenstates of the system, where $I_C$ is the classical limit
of $I_Q$, and $R_Q$ has no classical analog.  These operator averages
resolve features at significantly smaller scale than naively expected
from considering just the Floquet eigenstates alone. 

Further, both the classical and quantum geometries reflect the symmetry 
properties of the underlying Hamiltonian. That is, both are organized 
around phase-space points of high symmetry~\cite{haake}. Classically 
these are the stable and unstable classical periodic orbits. Since 
$I_Q$ and $R_Q$ also reflect the location of phase-space points of 
high symmetry, we obtain the observed correlations between classical
and quantum behavior.
We emphasize that our empirical results on the correlation between
quantum and classical measures at small $j$ make it necessarily true 
that the semiclassical perspective cannot apply; we develop
a new explanation for the correlation we see, and we consider the possible
applicability to higher $j$ as well. 
However, the higher $j$ regime is not the central focus of this paper.

In short, we show a strong correlation between classical dynamics and
quantum entanglement. In contrast to the standard understanding we
find that (a) this exists for the extreme quantum limit of a two-qubit
system, where the correspondence principle cannot be evoked, (b)
persists in the absence of chaos, and (c) is visible via single
trajectory measures.  We argue that this is (d) due to symmetry
considerations alone. We also see other interesting features of the
quantum entanglement dynamics not previously considered, specifically
that these dynamics are demonstrably periodic or quasi-periodic as a
function of number of kicks and $\kappa$. We discuss these issues in
detail below, starting with a short introduction to the kicked top
Hamiltonian.

\def \scaleval {.12} 

\begin{figure*}  
  \centering
  \begin{tabular}{c c c}
    \includegraphics[scale=\scaleval]{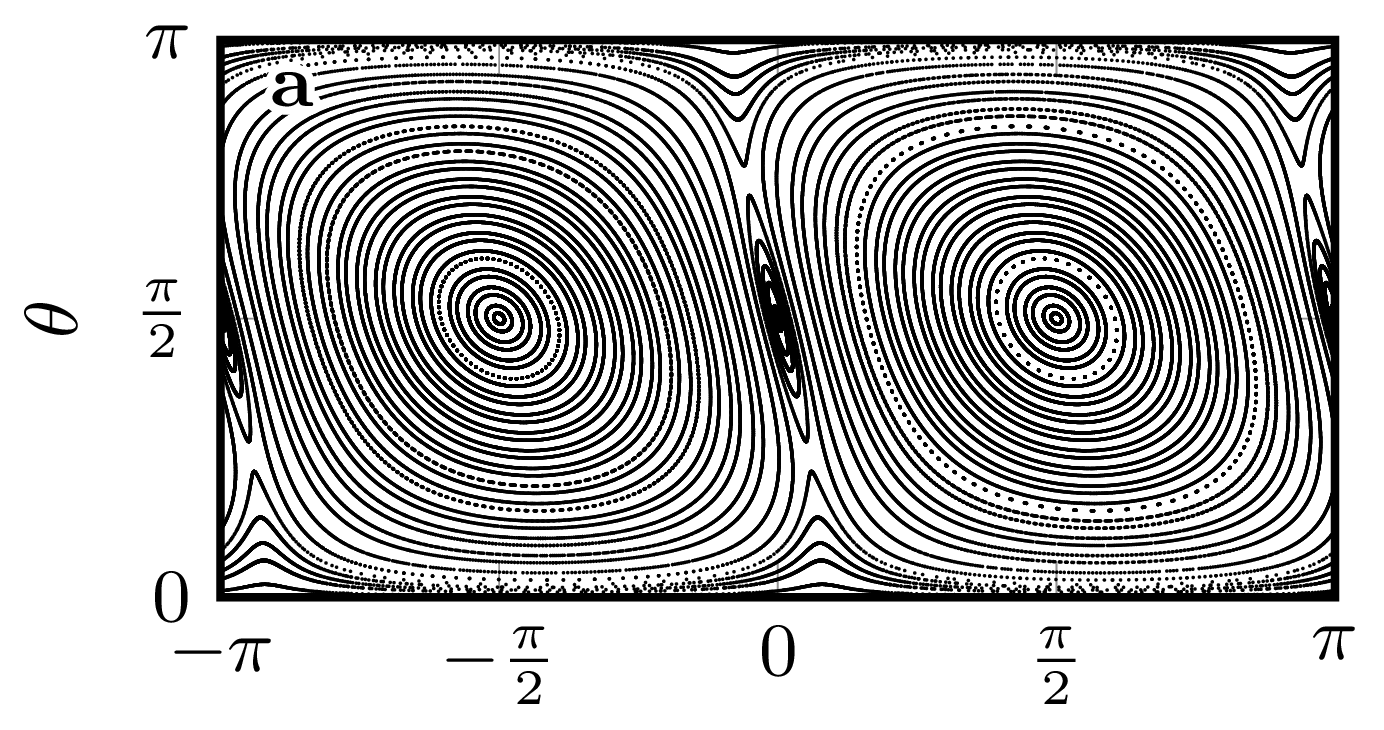}
    &
      \includegraphics[scale=\scaleval]{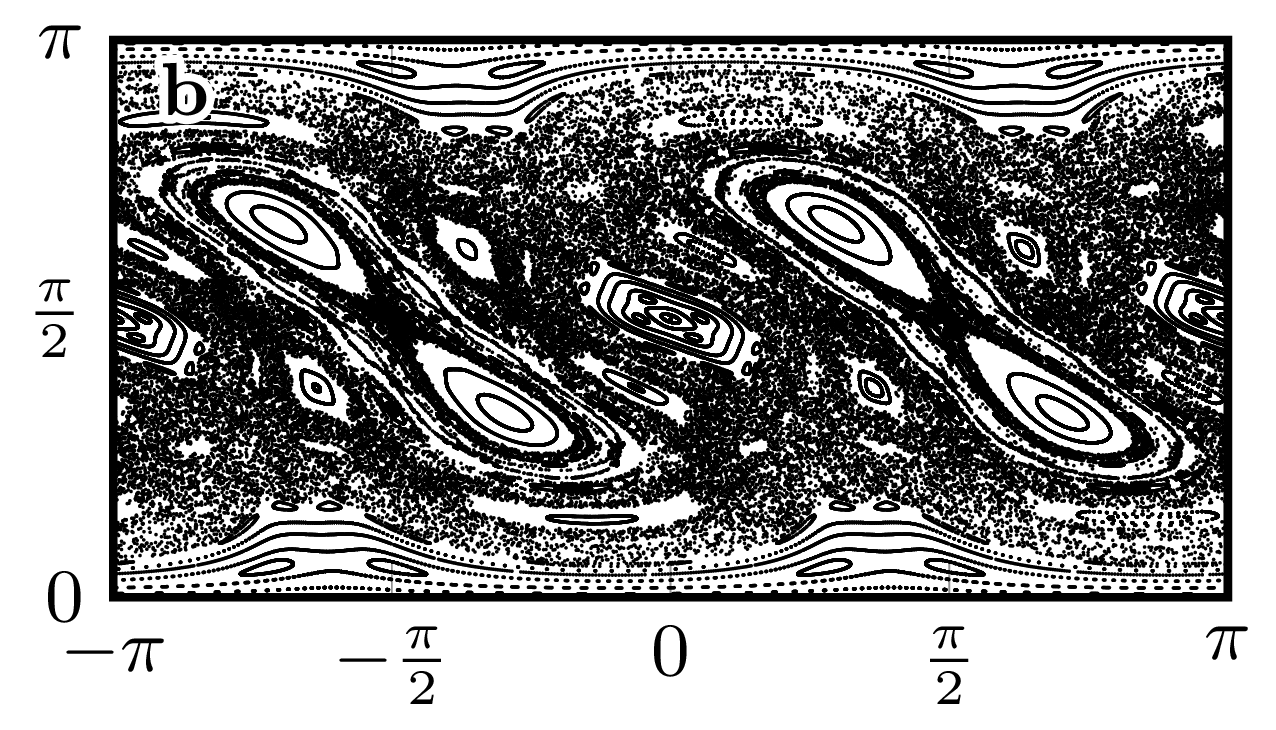}
    &
      \includegraphics[scale=\scaleval]{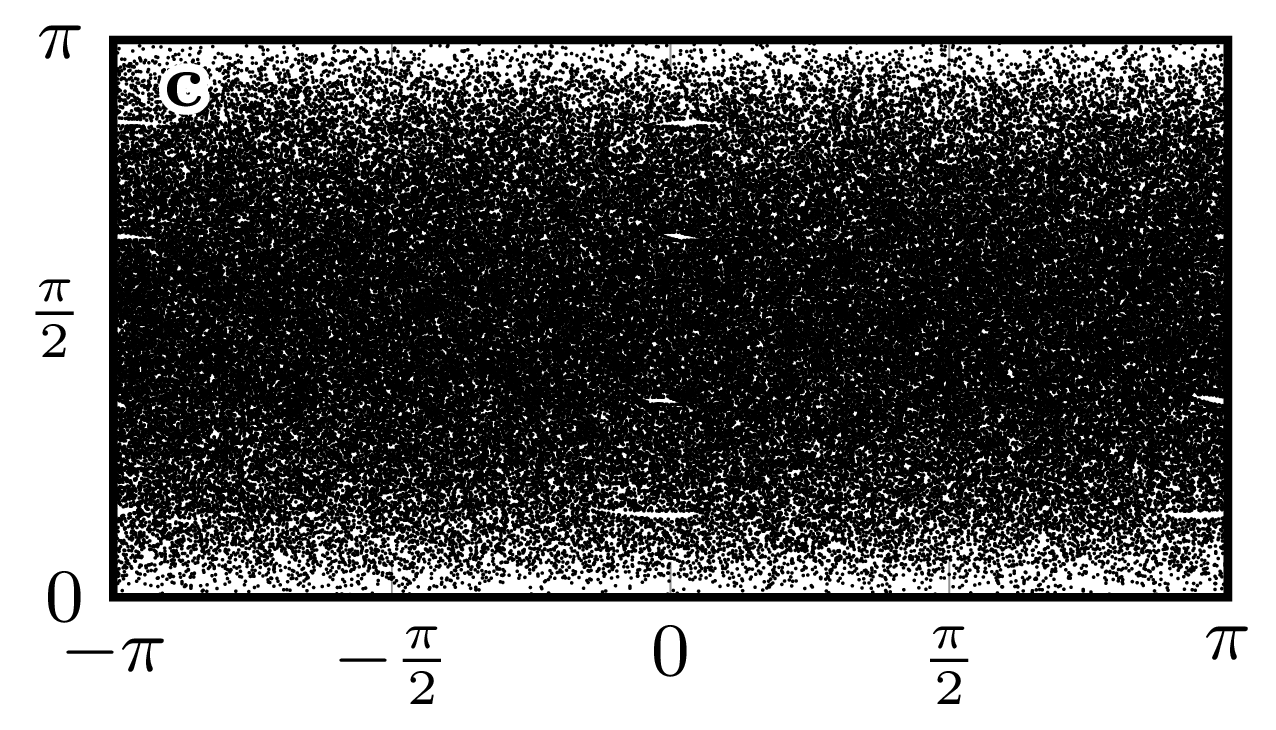}
    \\
    \includegraphics[scale=\scaleval]{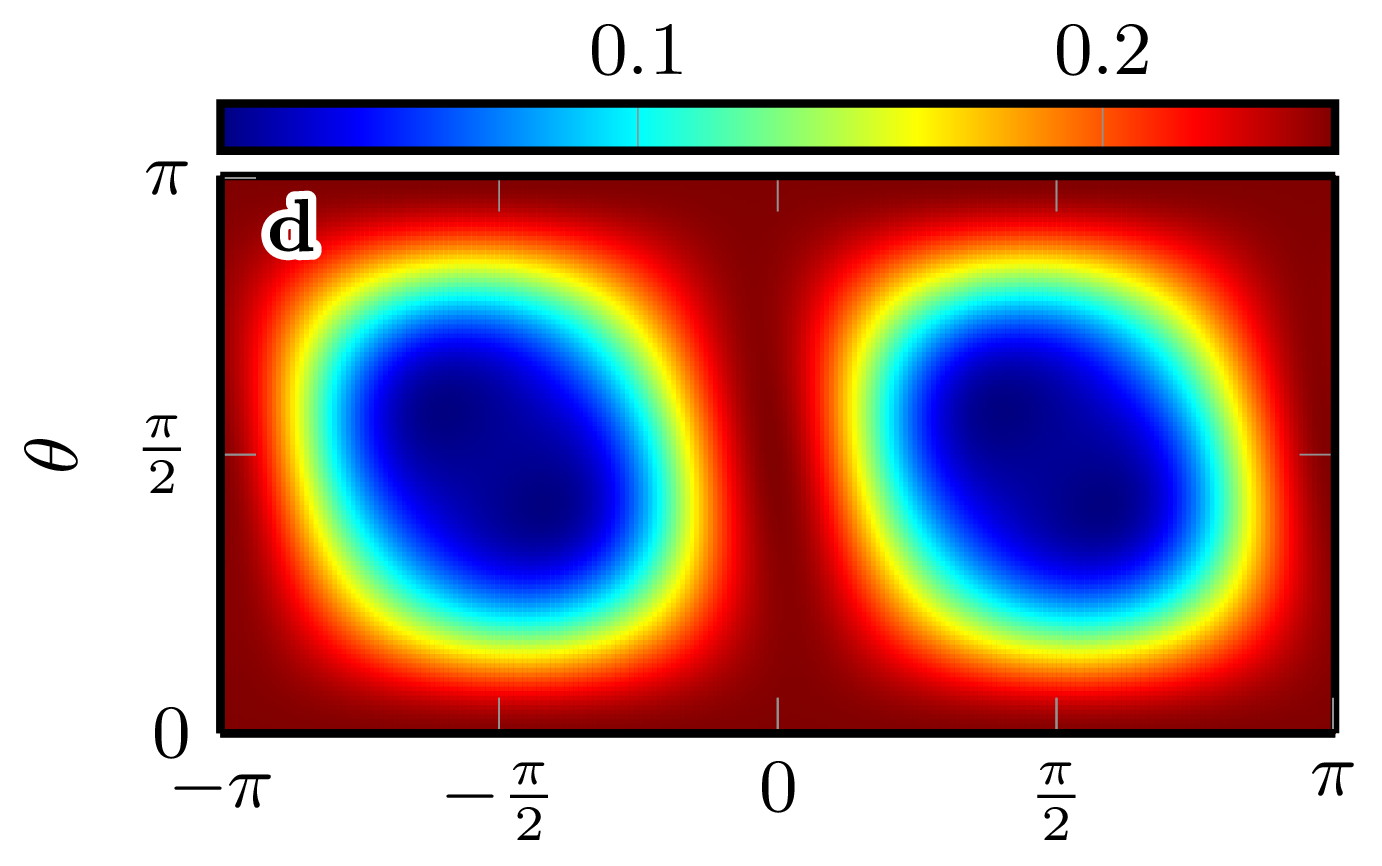}
    &
      \hspace{3pt}
      \includegraphics[scale=\scaleval]{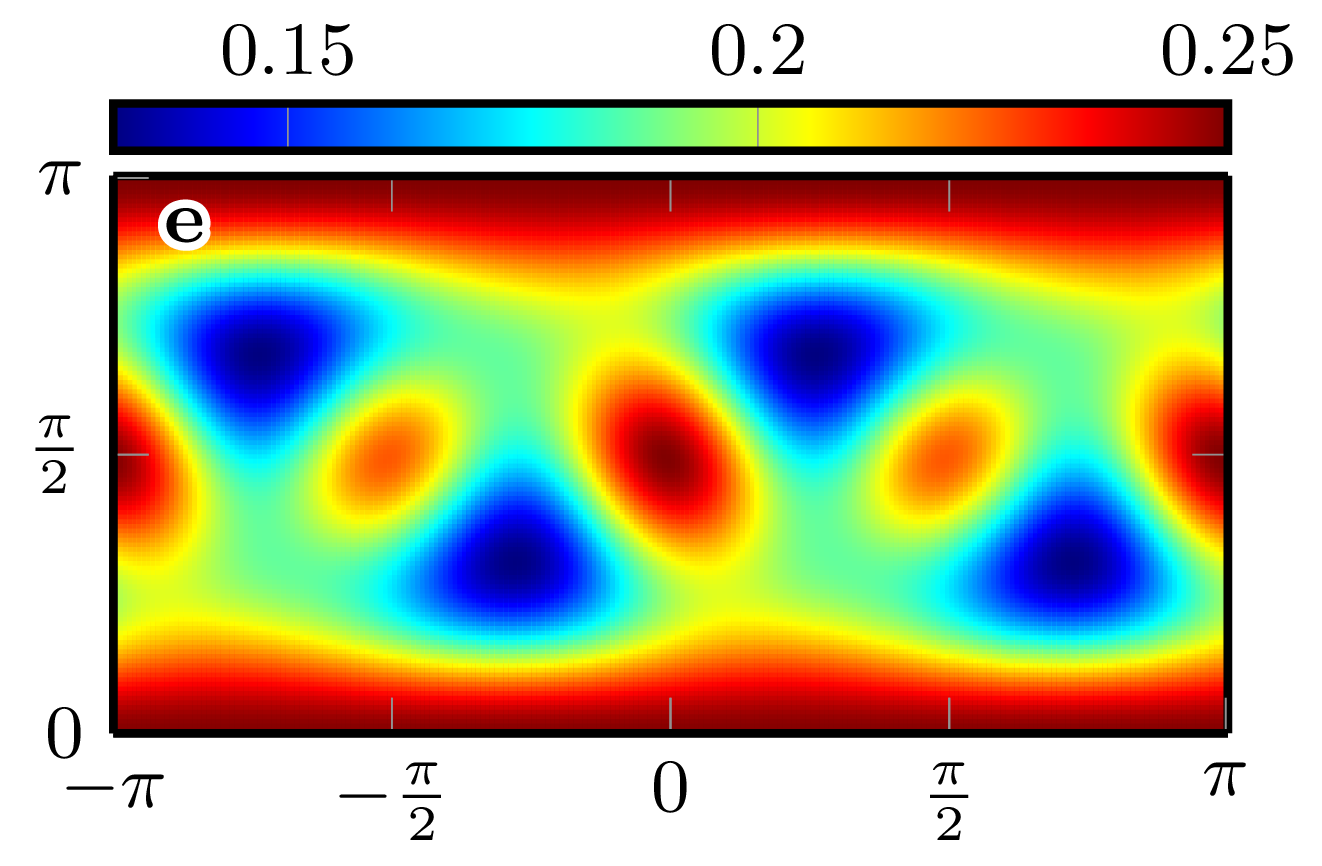}
    &
      \includegraphics[scale=\scaleval]{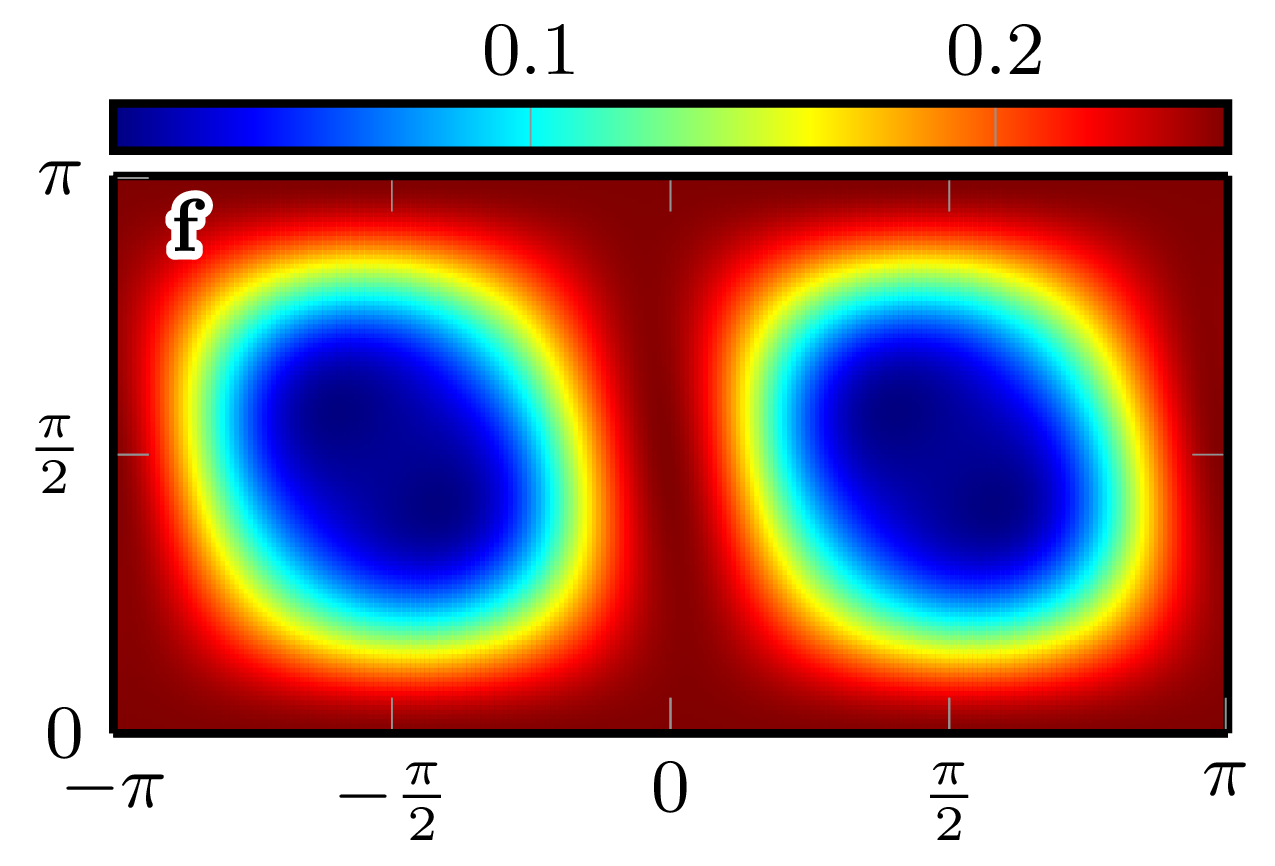}
    \\
    \includegraphics[scale=\scaleval]{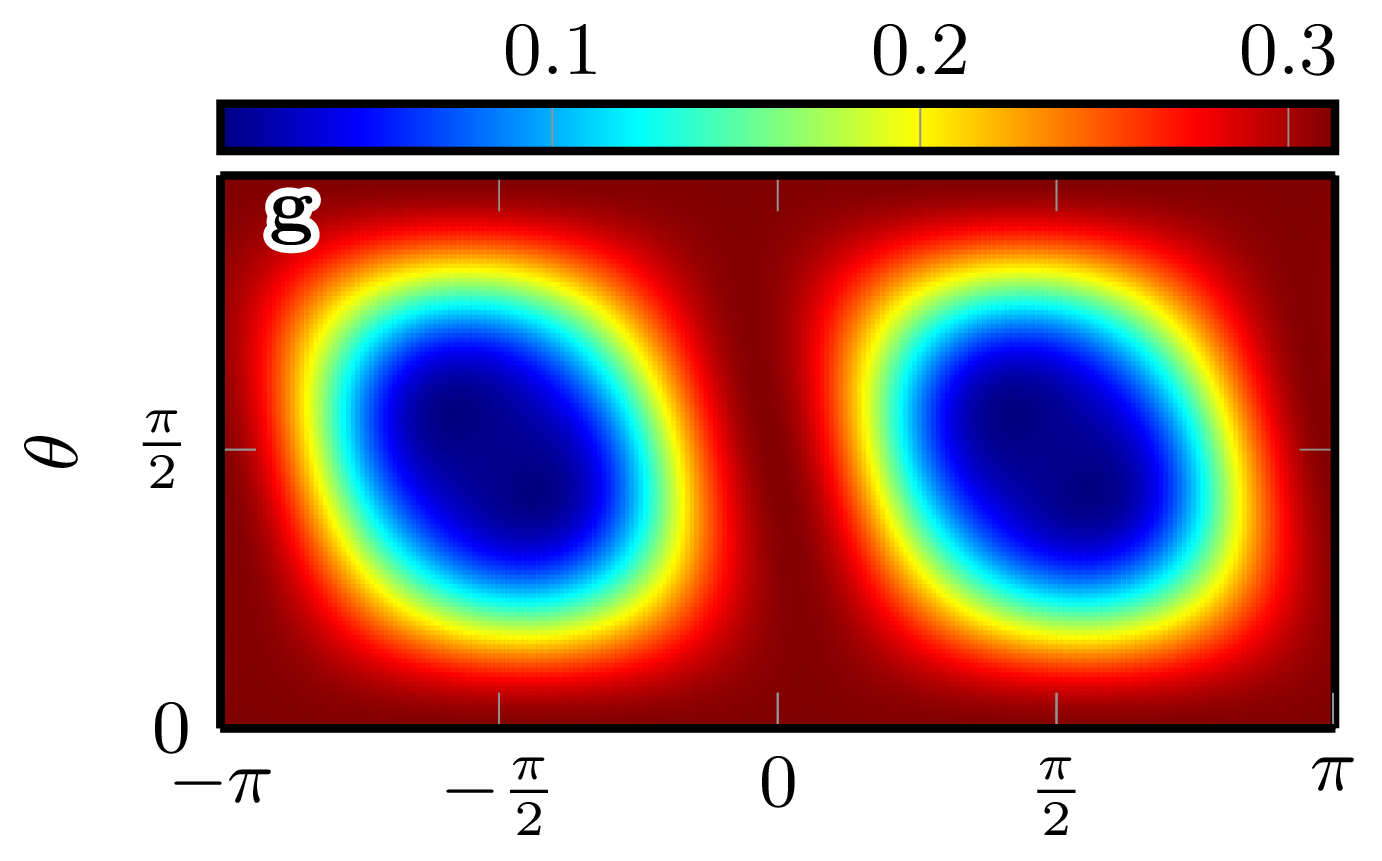}
    &
      \includegraphics[scale=\scaleval]{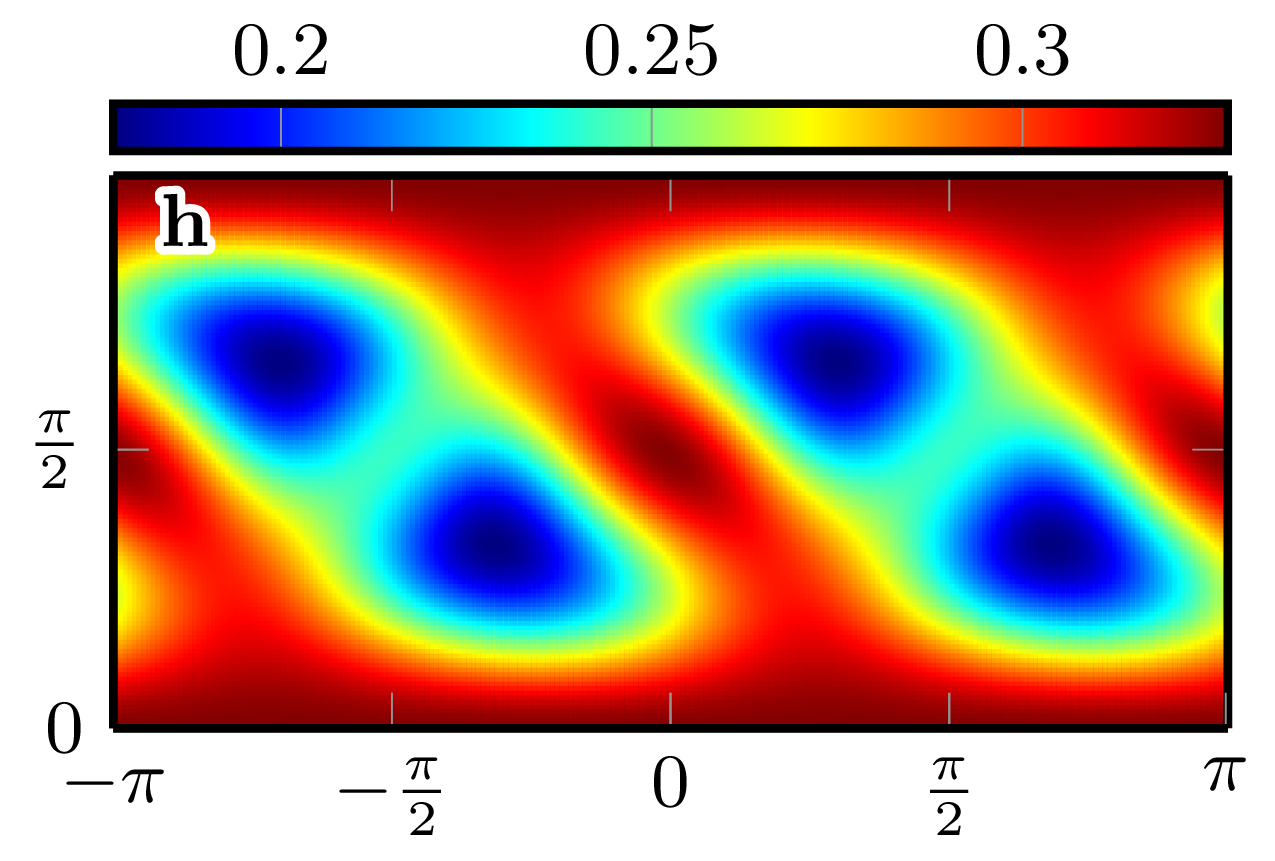}
    &
      \includegraphics[scale=\scaleval]{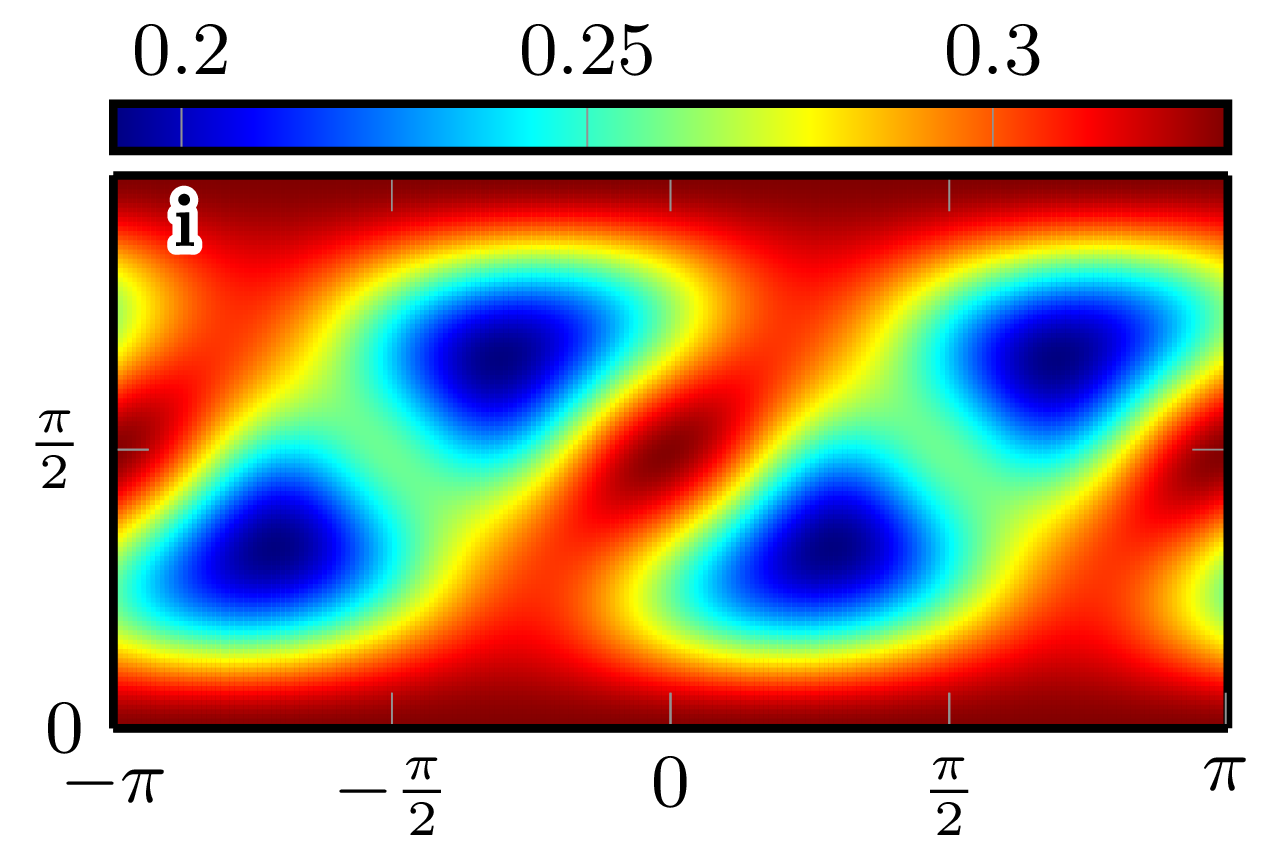}
    \\
    \includegraphics[scale=\scaleval]{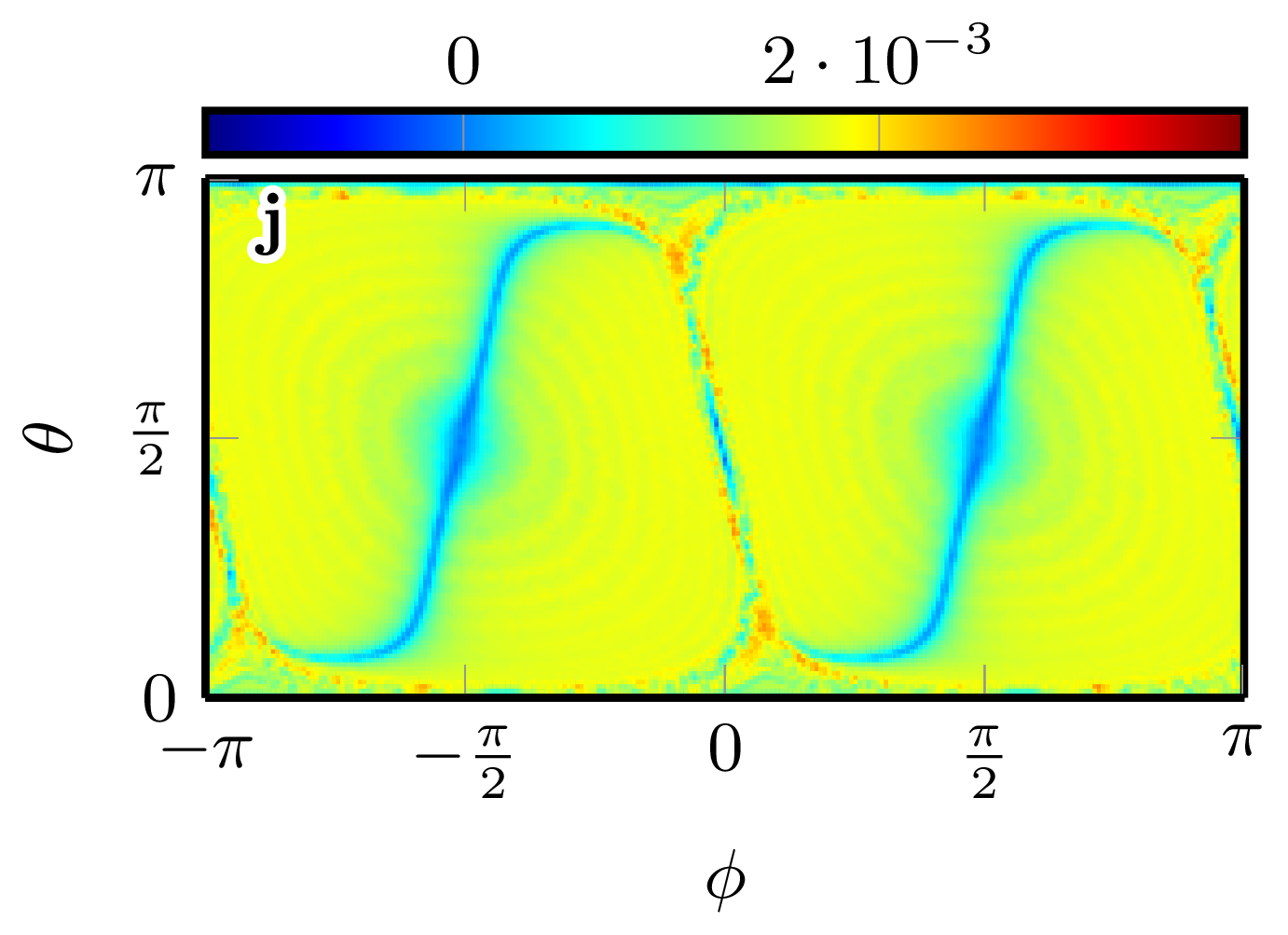}
    &
      \includegraphics[scale=\scaleval]{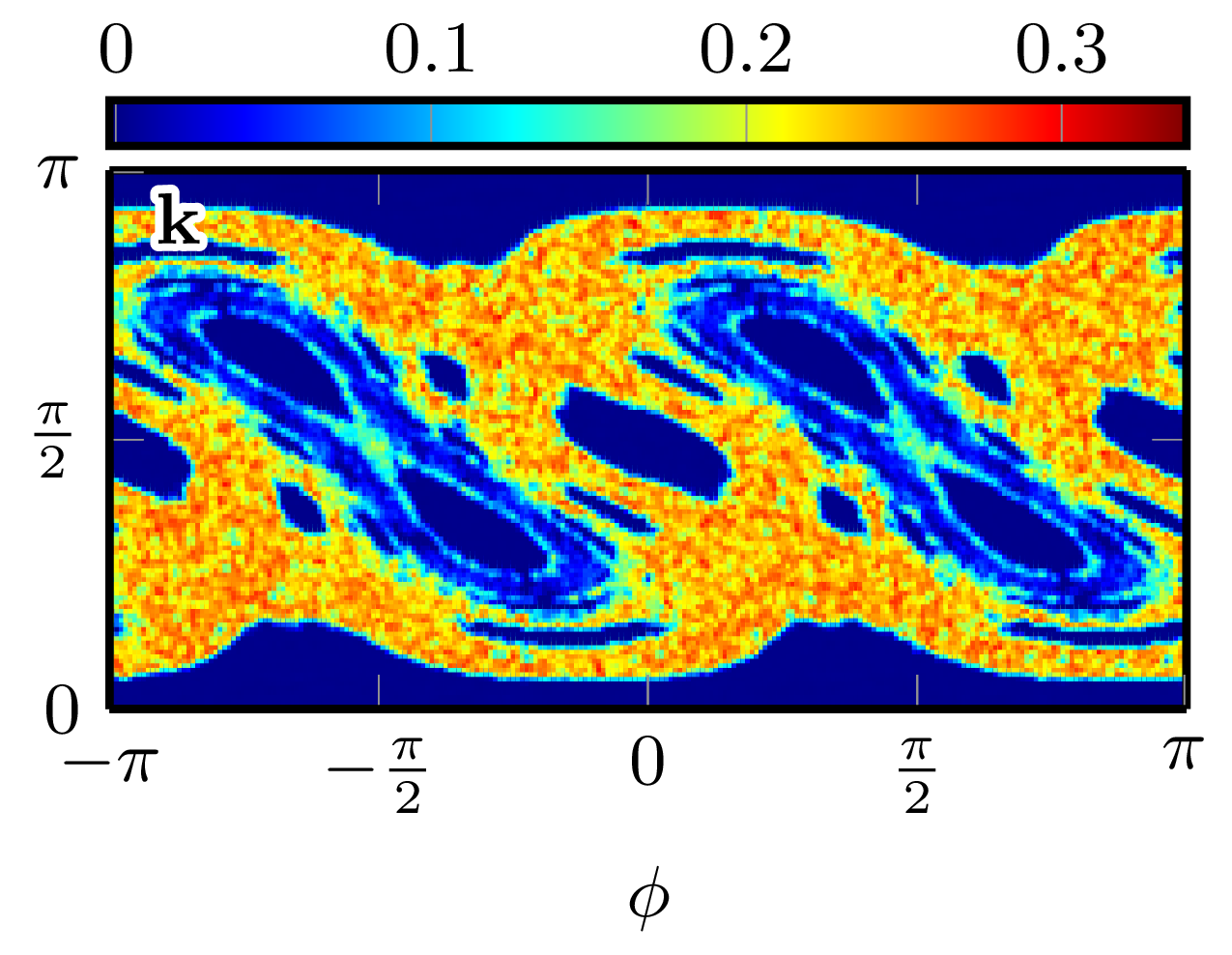}
    &
      \includegraphics[scale=\scaleval]{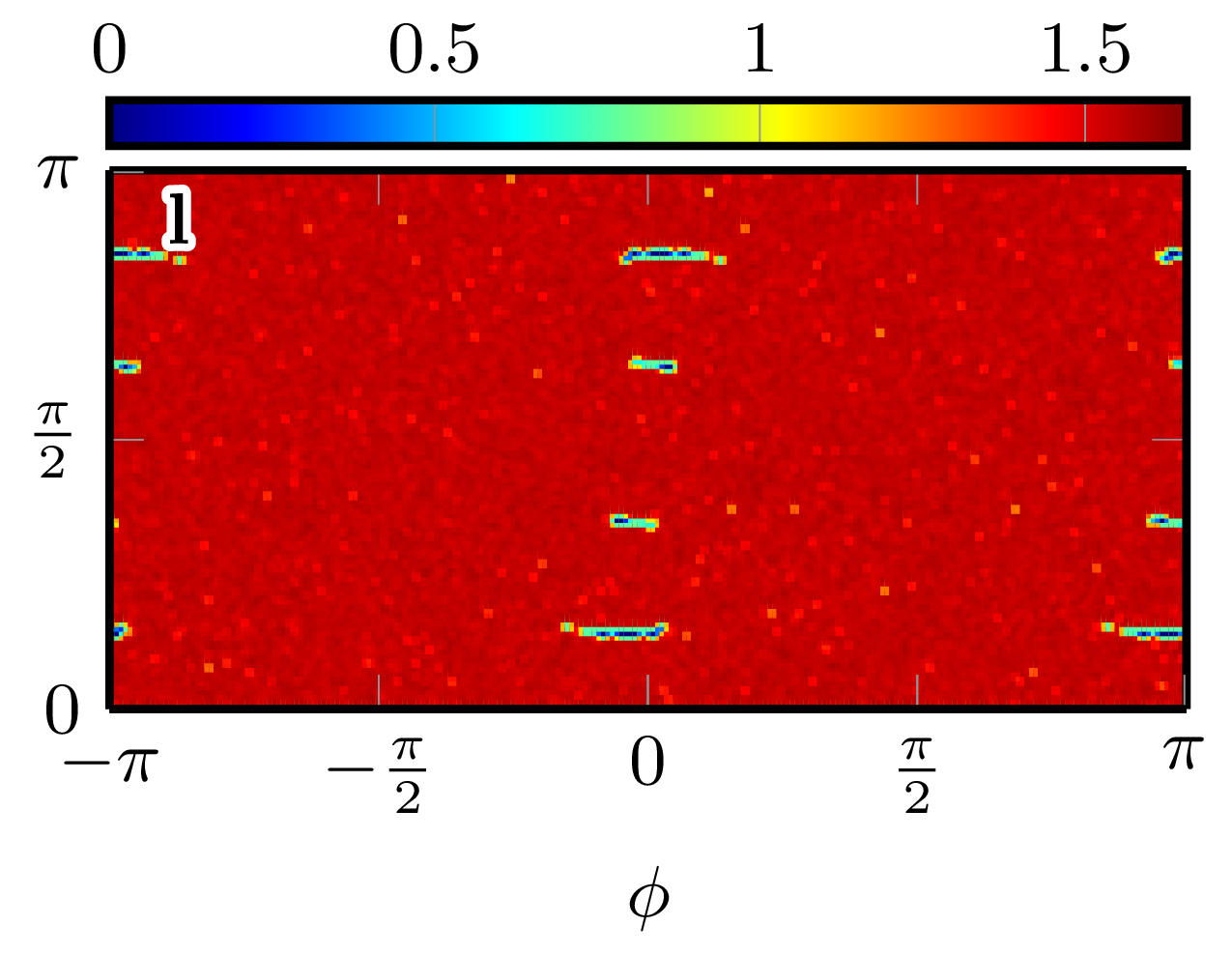}
  \end{tabular}
  \caption{
(a) Classical Poincar\'e section for $\kappa=0.5$. 
(b) Same but for $\kappa=2.5$. 
(c) Same but for $\kappa=2\pi + 0.5$.
(d) Infinite-time averaged entanglement $S_Q$ as a function of initial
condition for two qubits for
$\kappa=0.5$.
(e) Same but for $\kappa=2.5$.
(f) Same but for $\kappa=2\pi + 0.5$.
(g) Numerically computed (over 200 kicks) average entanglement $S_Q$ as
a function of initial condition for three qubits for $\kappa=0.5$.
(h) Same but for $\kappa=2.5$.
(i) Same but for $\kappa=2\pi + 0.5$.
(j) Finite time Lyapunov exponent (calculated over 2600 kicks) as a
function of initial condition for $\kappa=0.5$.
(k) Same but for $\kappa=2.5$.
(l) Same but for $\kappa=2\pi + 0.5$.
  \label{fig:bigfig}
    Note that these are images of a sphere projected
    onto the plane, so that the left and right edges are connected and
    area is distorted. In particular, the polar regions around
    $\theta=0$ and $\theta=\pi$ are the same size and shape as the
    regions around $(\theta,\phi)=(\pi/2,0)$ and $(\pi/2,\pi)$. Also
    note the different scales for each plot.}

\end{figure*}

\section{Background: The Kicked Top}
\label{sec:backgr-kick-top}

The kicked top is a spin $\vec J$ evolving under the Hamiltonian
\begin{equation}
  \label{eq:hamiltonian}
  H = \hbar \frac{\pi}{2\tau} 
  \hat J_y + \hbar \frac{\kappa}{2j} 
  \hat J_z^2 \sum_{n=-\infty}^{\infty} \delta(t - n\tau),
\end{equation}
which describes the precession of $\vec{J}$ around the $y$ axis
combined with a periodic shearing kick around the $z$ axis. $\kappa$
parametrizes the strength (nonlinearity) of the kick, while $\tau$
is the time between kicks. We pay attention to the system only
immediately following each kick at times $T_n = n\tau$, 
and thus obtain a map (understood as the Poincar\'e map of the Hamiltonian
flow), and a discrete unit of time $n$ defined by the number of kicks 
that have occurred.  

Using this discrete time description, the quantum system is most
conveniently studied via the Floquet 
operator~\cite{haake}
\begin{equation}
  \label{eq:floquet}
  \hat U = \exp(-i\frac{\kappa}{2j}\hat{J_z^2})\exp(-i\frac{\pi}{2}\hat J_y).
\end{equation}
If we expand an initial state $|\psi(0)\rangle$ in terms of
eigenvalues $\{\xi_i\}$ and eigenvectors \cite{fn1}
$\{|\xi_i\rangle\}$ of $\hat U$, then we can write the state at time
$n$ as
\begin{equation}
  \label{eq:2}
  \ket{\psi(n)} = \hat U^n \ket{\psi(0)}
  =\sum_i \bra{\xi_i}\ket{\psi(0)}^n\ket{\xi_i}.
\end{equation}
As is standard, to allow for meaningful comparison with the classical
limit, we restrict our attention to the behavior of states which are
initially spin-coherent states, which are minimum uncertainty states
for spin systems. These states are generated from the angular momentum
eigenstate $\ket{j,j}$
\begin{equation}
\ket{\psi(0)}=\ket{\theta,\phi} = \hat R(\theta,\phi)\ket{j,j}.
\end{equation}
Here the labels $j$ indicate eigenvalues for $\hat J^2$ and
$\hat J_z$, and the rotation $\hat R$ is defined as
\begin{equation}
  \hat R(\theta,\phi) = \exp\big[i\theta(\hat J_x\sin \phi
  - \hat J_y\cos\phi)\big]\label{eq:coherentstate}
\end{equation}
where $\phi\in [-\pi,\pi),\ \theta \in [0,\pi)$. These spin-coherent
states are thus centered at some location on the sphere $(\theta,\phi)$,
and these locations are the classical initial conditions against which 
their behavior is to be compared.

Following previous studies, we study the entanglement of the system by
considering the situation where the spin $J$ is composed of $2j$
spin-1/2 particles, or qubits, such that
$\vec{J} = \sum_{i=1}^{2j}\vec{s}_i$ for individual spins
$\vec{s}_i$. While several different types of entanglement measures
can be considered, all of them have been shown to have essentially the
same broad behavior. We focus on the measure most studied, the
bipartite entanglement between any one of the qubits and the subsystem
made up of the remaining $2j-1$ qubits. This entanglement is
quantified by computing the linear entropy
\begin{equation}
  \label{eq:3}
  S = 1 - \rm{Tr}\rho_1^2,
\end{equation}
where $\rho_1$ denotes the density operator for any one of the qubits,
obtained by taking the partial trace over the $2j-1$ other qubits. 
Since the dynamics are restricted to the symmetric subspace of the 
total spin, the entropy can also be written as~\cite{ghosesilberfarb}
\begin{equation}
S = \frac{1}{2}\bigg[1 - \frac{1}{j^2} \big(\langle \hat
J_x\rangle^2+\langle \hat J_y\rangle^2+\langle \hat
J_z\rangle^2\big)\bigg].\label{eq:entropy1}
\end{equation}

\begin{figure*}
  \centering

  \begin{tabular}{c c c}
    \includegraphics[scale=\scaleval]{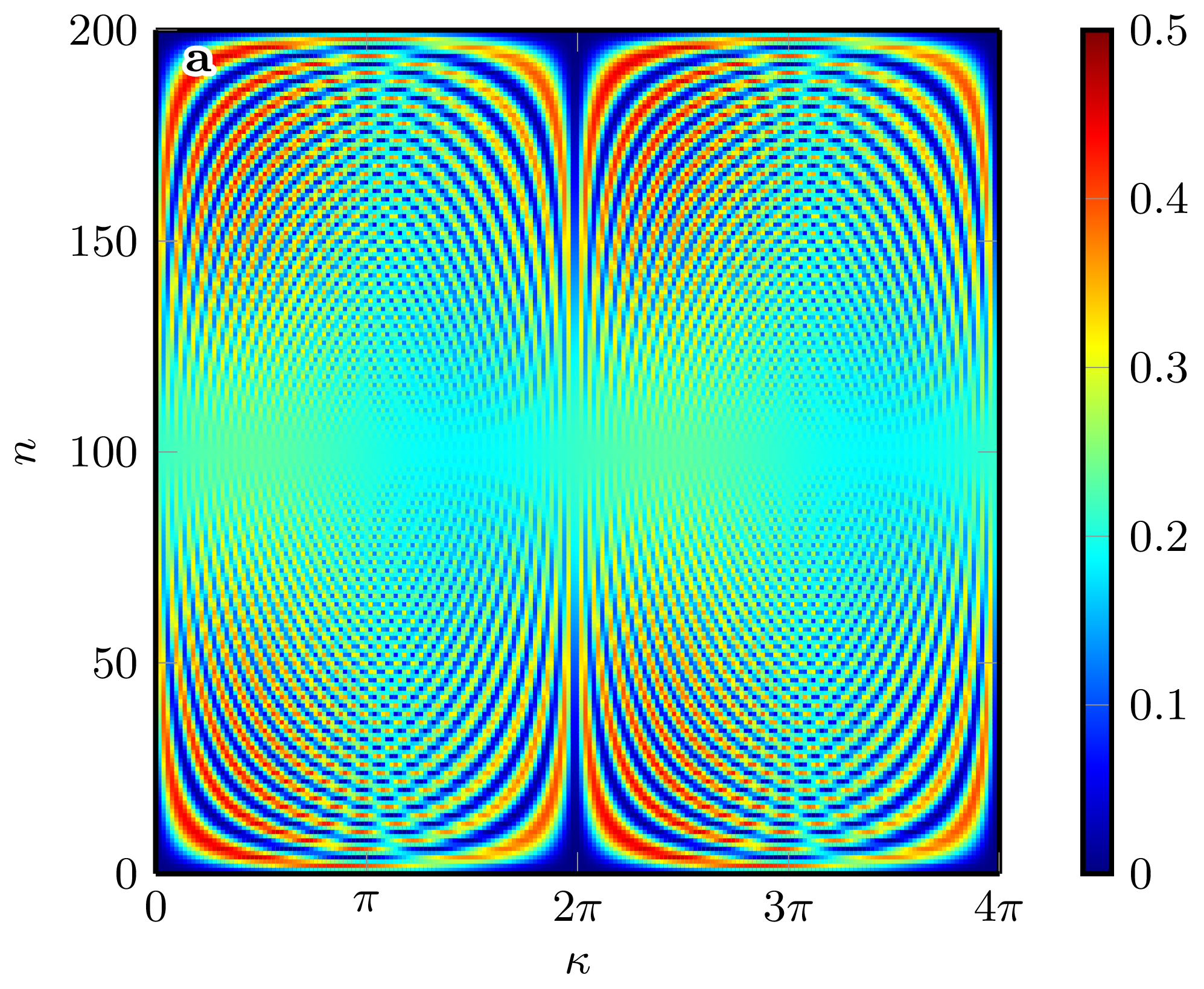}
    &\hspace{.5cm}
    &\includegraphics[scale=\scaleval]{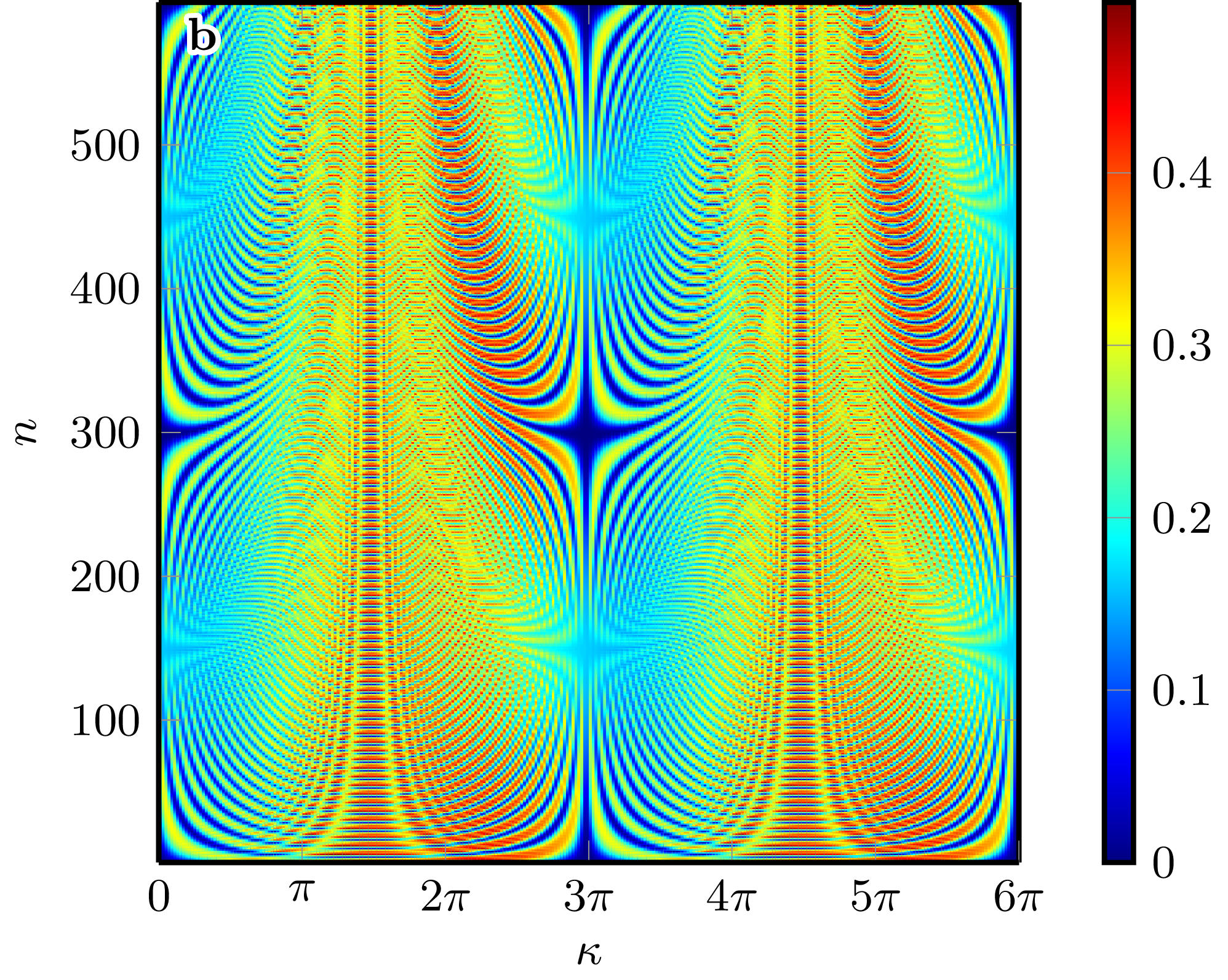}
  \end{tabular}
  \caption{\label{F:butterflies} (a) Entanglement entropy $S$ plotted as a
    color (color-coding shown in the bar on the right of both
    figures) for a two qubit system ($j=1$). It is plotted as a function of 
    time $n$ on the vertical axis and kick strength $\kappa$ on the 
    horizontal axis for an arbitrarily chosen initial condition 
    $(\theta,\phi)=(1.2,0.3)$. 
    (b) The same, computed numerically for three qubits ($j=3/2$) with
    initial condition $(2.5,1.1)$. Note the (quasi-)periodicity in
    both $\kappa$ and $n$, and the longer periods in both $\kappa$ and
    $n$ for three qubits compared to two qubits.}
\end{figure*}

In the classical limit $j\to\infty$, the system is described by the 
point dynamics of an angular momentum vector which we describe by its
coordinates $(x,y,z)$. Reference~\cite{haake} gives the classical map $F$
from time-step $n$ to $n+1$ for this vector:
\begin{eqnarray}
  x_{n+1} & = & z_n \cos(\kappa x_n) + y_n \sin(\kappa x_n),  \\
  y_{n+1} & = & -z_n \sin(\kappa x_n) + y_n \cos(\kappa x_n), \\
  z_{n+1} & = & - x_n.
\end{eqnarray}
Since total angular momentum is conserved~\cite{haake}, these dynamics
occur on the surface of a sphere of unit radius, with the usual
relation
$(x,y,z)= (\sin\theta\cos\phi, \sin\theta\sin\phi,\cos\theta)$. The
restriction to the surface of the sphere means that there are
effectively two phase-space variables $(\theta,\phi)$.  For low
$\kappa$ all initial conditions in the phase space show regular
behavior. Around $\kappa \approx 1.0$, chaos emerges for certain
initial conditions near unstable fixed points. The effect of
increasing $\kappa$ beyond this value is to increase both (a) the
extent of phase space displaying chaotic behavior (the number of
initial conditions displaying chaos) and (b) the degree of chaos---the
rapidity with which initially infinitesimally close initial conditions
separate in their trajectories.  The behavior of a set of initial
conditions is shown for three different values of $\kappa$ in the
first row ({\bf a-c}) of Fig.~\ref{fig:bigfig}. We quantify the
chaotic behavior in detail later when comparing the behavior with the
quantum entanglement dynamics.

\section{Entanglement dynamics of the 2-qubit kicked top}
\label{sec:entangl-dynam-2}

One of the advantages of working at small $j$ is that for the smallest
non-trivial quantum system ($j=1$), we can carry out many calculations
analytically that must be considered numerically even for $j=3/2$. In
particular, we find a closed-form solution for the entanglement
$S(\theta,\phi,\kappa,n)$ as a function of initial position, kick
strength, and time; this also allows an expression of its
infinite-time average. If we examine the quantum `orbit' of the vector
$\expval*{\vec J}=(\expval*{\hat J_x},\expval*{\hat J_y},\expval*{\hat
  J_z})$, we can gain insight into the entropy as expressed in
Eq.~\ref{eq:entropy1}, since the length of this vector is the quantity
of interest. The orbits are most easily understood by splitting them
up into even and odd time-steps, for kicking strength $\kappa$ and
initial condition $\ket{\theta,\phi}$:
\begin{align}
  \label{eq:5}
  \expval*{\hat J_x(n)} &=
  \begin{cases}
    (-1)^{n/2}\big[\sin\theta\cos\phi\cos(\frac{\kappa}{2}\frac{n}{2})\\
    - \sin\theta\cos\theta\sin\phi\sin(\frac{\kappa}{2}\frac{n}{2})\big]
    & n\text{ is even}\vspace{.4cm}\\
    (-1)^{(n+1)/2}\big[\cos\theta\cos(\frac{\kappa}{2}\frac{n+1}{2})\\
    +\cos\phi\sin\phi\sin^2\theta\sin(\frac{\kappa}{2}\frac{n+1}{2})\big]
    & n\text{ is odd}
  \end{cases}\\
  \expval*{\hat J_y(n)} &=
  \begin{cases}
    \sin\theta\sin\phi
    & n\text{ is even}\vspace{.1cm}\\
    \sin\theta(\sin\phi\cos\frac{\kappa}{2}-\cos\theta\cos\phi\sin\frac{\kappa}{2})
    & n\text{ is odd}
  \end{cases}\\
  \expval*{\hat J_z(n)} &= -\expval*{\hat J_x(n-1)}
\end{align}

The expression for $\expval*{\hat J_y(n)}$ does not depend on $n$
beyond its parity, which causes the orbits to lie in planes parallel
to the $xz$ plane in the shape of two deformed ellipses (at even $n$
it is exactly an ellipse). When $\kappa$ is an irrational fraction of
$\pi$, the dynamics are quasiperiodic in $n$ and this orbit is
explored ergodically. If $\kappa$ is a rational fraction of $\pi$,
then the dynamics are periodic and a finite subset of the orbit is
explored. This (quasi-)periodicity holds for higher values of $j$, as
can be seen from the form of the Floquet operator
(Eq.~\ref{eq:floquet}) and the resulting eigenvalues. In general, the
length of the period increases with increasing $j$, which is why this
periodicity has not been observed in previous studies that focus on
higher $j$.

These results allow an explicit evaluation of $S$, which we plot in
Fig.~\ref{F:butterflies} as a function of both $\kappa$ and $n$ for an
arbitrary initial condition $\ket{\theta,\phi}$; the
(quasi-)periodicity is clearly visible in these `butterfly wing'
plots. We can also take an infinite-time average
$S_Q(\theta,\phi,\kappa)=\overline{S(\theta,\phi,n,\kappa)}$ that
relies on the ergodic exploration of these orbits in the quasiperiodic
case. We show $S_Q$ for a selection of $\kappa$ values in the second
row ({\bf d-f}) of Fig.~\ref{fig:bigfig}. $S_Q$ is also periodic in
$\kappa$, and in fact the Floquet expansion shows that any finite-$j$
system is also periodic in $\kappa$ where this period increases with
increasing $j$.  For example, the Floquet eigenvalues of the $two$-qubit
system are
$\{\me^{-i \kappa /2},-i \me^{-i \kappa /4},i \me^{-i \kappa /4}\}$,
so the dynamics are unchanged if $\kappa \to \kappa + 8m\pi$ for any
integer $m$. In particular, since there is no entanglement at all at
$\kappa=0$, there is also no entanglement for $\kappa = 8m\pi$. In
fact, due to averaging effects, $S_Q$ has period $2\pi$ in $\kappa$
which is confirmed by comparing the first and third columns of
Fig.~\ref{fig:bigfig}.  We can also calculate the average entanglement
for three qubits numerically, as shown in the third row ({\bf g-i}) of
Fig.~\ref{fig:bigfig}. This calculation is necessarily a finite-time
average, but it allows comparison with recent experiment~\cite{pedram}
and shows that our observations are not unique to the two-qubit case.

As an alternative, for any number of qubits, $S_Q$ can be written in
terms of the Floquet eigenbasis (with
$C_{k}=\langle\xi_k|\theta,\phi\rangle$) as
\begin{align} \label{E:entropy}
  S_Q&=
  \frac{1}{2}-\frac{1}{2j^{2}}\overline{\sum_{i} \Big[
  \langle \psi(n) | \hat J_{i} | \psi(n) \rangle^{2}\Big]} \nonumber \\
  &=\frac{1}{2}-\frac{1}{2j^{2}}\overline{\sum_{i}
  \Big[\sum_{k,l}\left(\xi_{k}^*\xi_{l}\right)^{n}C_{k}^*C_{l}
  \langle\xi_{k}|\hat J_{i}|\xi_{l}\rangle\Big]^{2} } \nonumber \\
  &=\frac{1}{2}-\frac{1}{2j^{2}}\sum_{k,l,p,q}\sum_{i}
   C_{k}^* C_{l}  C_{p}^* C_{q}
  \langle \xi_{k}|\hat J_{i}|\xi_{l}\rangle
  \langle \xi_{p}|\hat J_{i}|\xi_{q}\rangle \nonumber \\
  &\equiv \frac{1}{2}-\frac{1}{2j^{2}}\sum_{k,l,p,q}
   E(k,l,p,q) \nonumber \\
  &\text{\qquad for } \{k,l,p,q: \xi_{k}^*\xi_{l}\xi_{p}^*\xi_{q}=1\}
\end{align}
where the last line defines the $E(k,l,p,q)$ that we will discuss later.
This essentially separates the `AC' components from the `DC' components
(that is, the `AC' terms average to zero).
Since the only time-dependence is in the power of the product of the 
eigenvalues, which are all modulus 1 complex numbers, we have to satisfy
the condition $\xi_{k}^*\xi_{l}\xi_{p}^*\xi_{q}=1$ to find the
DC components. However, since this is an {\em exact} condition, it is 
not trivial to use this equation for numerical work. The 2-qubit case 
is the only one in which we can diagonalize $\hat U$ entirely analytically, 
and hence the only case in which we can perform a straightforward computation 
of the `DC' results. However, this expansion yields theoretical insight 
as we discuss in Sec.~\ref{sec:class-quant-ignor}.

\section{Classical chaos and quantum-classical similarities}
\label{sec:chaos-class-limit}

To proceed further in comparing the quantum and classical initial
condition dependence and behaviors, we start with quantifying the
degree of chaos in the classical kicked top. To do this we use the
(largest) Lyapunov exponent, which characterizes the time dependence
of how two orbits initialized close together in phase space diverge.
After choosing an initial point $(\theta,\phi)$, we evolve both the
map and the tangent vector to the map, rescaling the tangent vector to
a unit vector at each step. At each step the scale change in length is
recorded and then averaged. Formally, the Lyapunov exponent is
calculated as the average computed in the limit as $n\to\infty$. For
an ergodic system, the infinite-time Lyapunov exponent is independent
of initial condition. However for a generic classical Hamiltonian
system Lyapunov exponents depend on the initial condition. Studies of
this dependence, particularly as computed for finite-time Lyapunov
exponents and their time-dependent convergence have proved very useful
in characterizing the phase-space geometry of dynamical instability in
generic chaotic systems (see, for example, the discussion
in Ref.~\cite{prasad}). In this system we see regular, mixed 
(regular and chaotic regions co-existing in
phase-space), and completely chaotic behavior as $\kappa$
increases. We characterize these different behaviors by computing a
finite time Lyapunov exponent (using $n=2600$ and with transient
behavior rejected by discarding the first 100 steps) and mapping the
finite time Lyapunov exponent as a function of initial condition.

An example of how this finite time exponent is useful in ways the
infinite time exponent may not be is visible in the figure at
$\kappa=0.5$. Here, $\lambda$ is not identically zero, even though the
classical system is completely regular. All the deviations from zero
are very small, and these can be shown to converge to $0$ for
$t\to\infty$. However, these `slow-to-coverge' regions mark the
invariant manifolds of the fixed points of the classical dynamics. The
locations of these manifolds in fact determine the phase-space
separation into stability `islands' and `chaotic sea' as $\kappa$
increases. This underlines the idea (which we explore further below)
that the phase space is organized around the fixed points of the map
dynamics (simple periodic orbits of the flow).

\begin{figure}
  \includegraphics[scale=\scaleval]{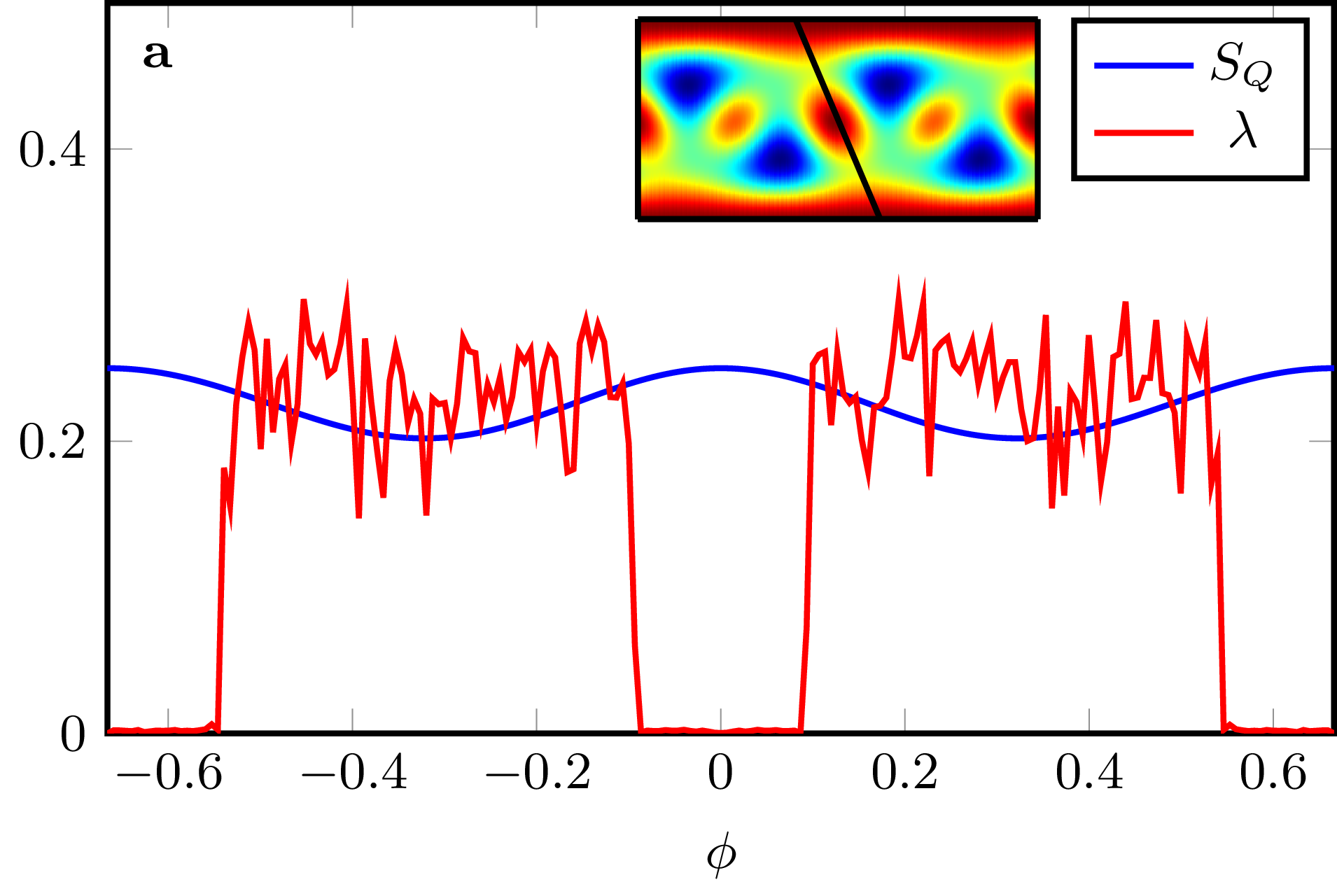}
  \includegraphics[scale=\scaleval]{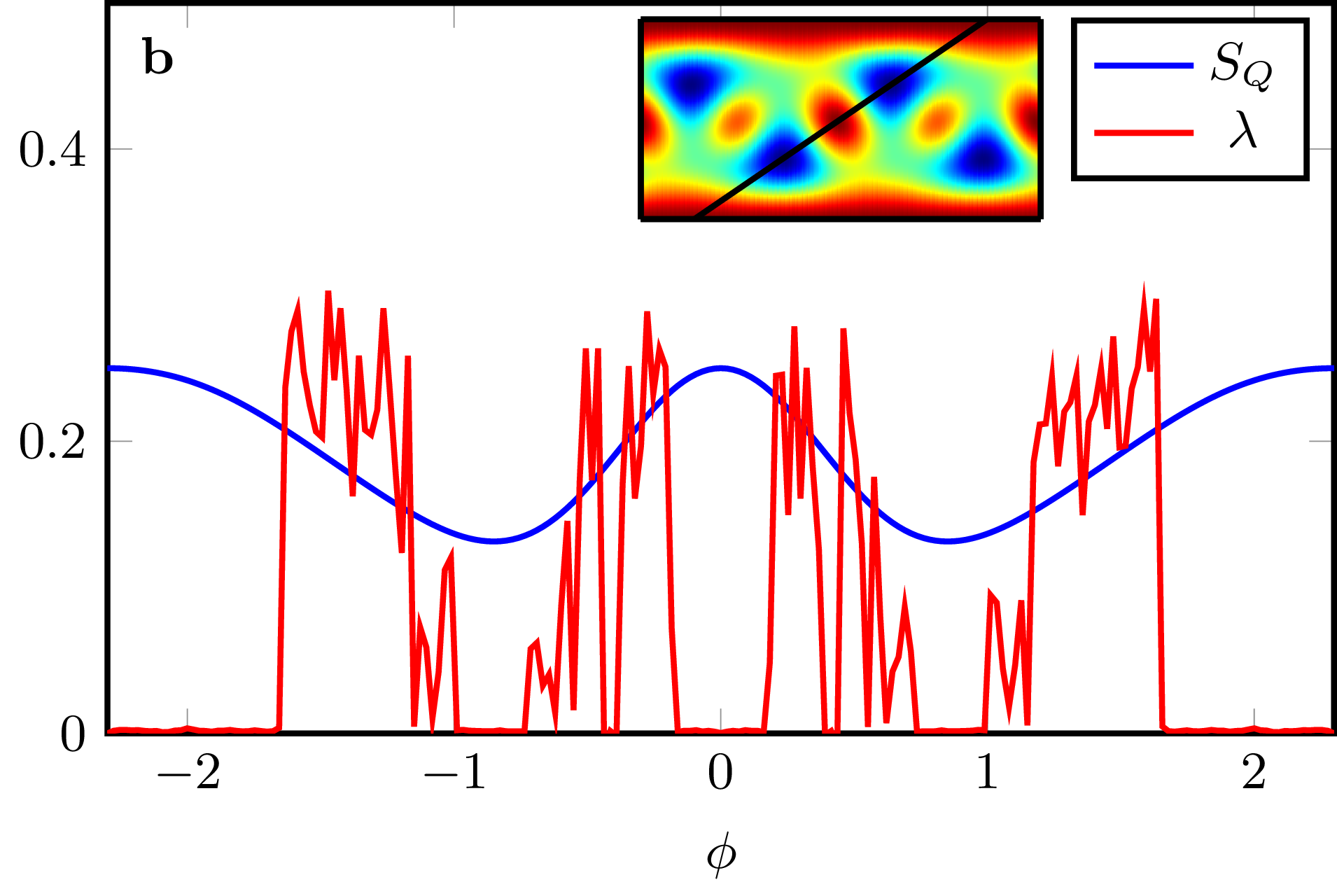}
  \caption{(a) Slices of the plots of time-averaged entanglement $S_Q$ and
    Lyapunov exponent $\lambda$ at $\kappa=2.5$. The slice taken is
    shown by the black line $\theta =(\pi/2)(\phi/\phi_0+1)$
    superimposed on the plot of $S_Q$ with $\phi_0=-0.666018$. 
    (b) Same except $\phi_0=2.29965$. Note the anti-correlation at the
    center and edges of both plots.}
  \label{fig:slices}
\end{figure}

With direct measures of both entanglement and chaos in hand, we now
compare the entanglement of the quantum system to the chaos in the
classical system as previously done in the literature. First, in
comparing the first and second rows of Fig.~\ref{fig:bigfig}, it is
immediately clear that the entanglement average has signatures of the
classical orbits.  That is, both classically and quantum mechanically,
different initial conditions lead to very different behavior, and the
boundaries between different behavior have approximately the same
location and shape in both figures. This would seem initially to
validate the previous consensus in the literature.  However, the fact
that this resemblance remains even in this extreme quantum limit means
that all previous arguments---which relied on the semiclassical nature
of the quantum system being studied---cannot hold.  Further, this
resemblance between the two different geometries exists without
classical chaos, as evidenced by the similarities between graphs of
the entanglement and the classical phase space (Figs.~\ref{fig:bigfig}
1{\bf d}--{\bf f} and 1{\bf a}--{\bf c}), respectively) at
$\kappa=0.5$. Finally, the plot of $\lambda$ for the classical system
at $\kappa =2.5$ (third row) shows that although the shapes of regions
around $(\theta,\phi)=(\frac{\pi}{2},0)$, $(\frac{\pi}{2},\pi)$,
$(0,0)$, and $(\pi,0)$ are similar to the shapes of regions in the
entanglement plot, these regions have a low Lyapunov exponent but high
entanglement. Both this anti-correlation and the similarity of region
boundaries are particularly visible in Fig.~\ref{fig:slices}, where
$S_Q$ and $\lambda$ are plotted together along a line of initial
conditions.

In general, we therefore see that both very high and very low levels
of entanglement are correlated with different types of regular
classical dynamics, but chaotic dynamics correspond to a level of
entanglement about halfway between these extremes~\cite{fn2}.

We also show a comparison at $\kappa=0.5+2\pi$. Here the quantum
entanglement geometry is clearly different from the classical
phase-space geometry. This clear break in the similarities of the
systems can be easily understood as arising from the fact that the
quantum system is periodic in $\kappa$ while classical phase space
becomes increasingly chaotic as $\kappa$ increases.

\subsection{Classical and quantum `ignorance' (delocalization) }
\label{sec:class-quant-ignor}

\begin{figure*}
  \centering
  \begin{tabular}{c c c}
    \includegraphics[scale=\scaleval]{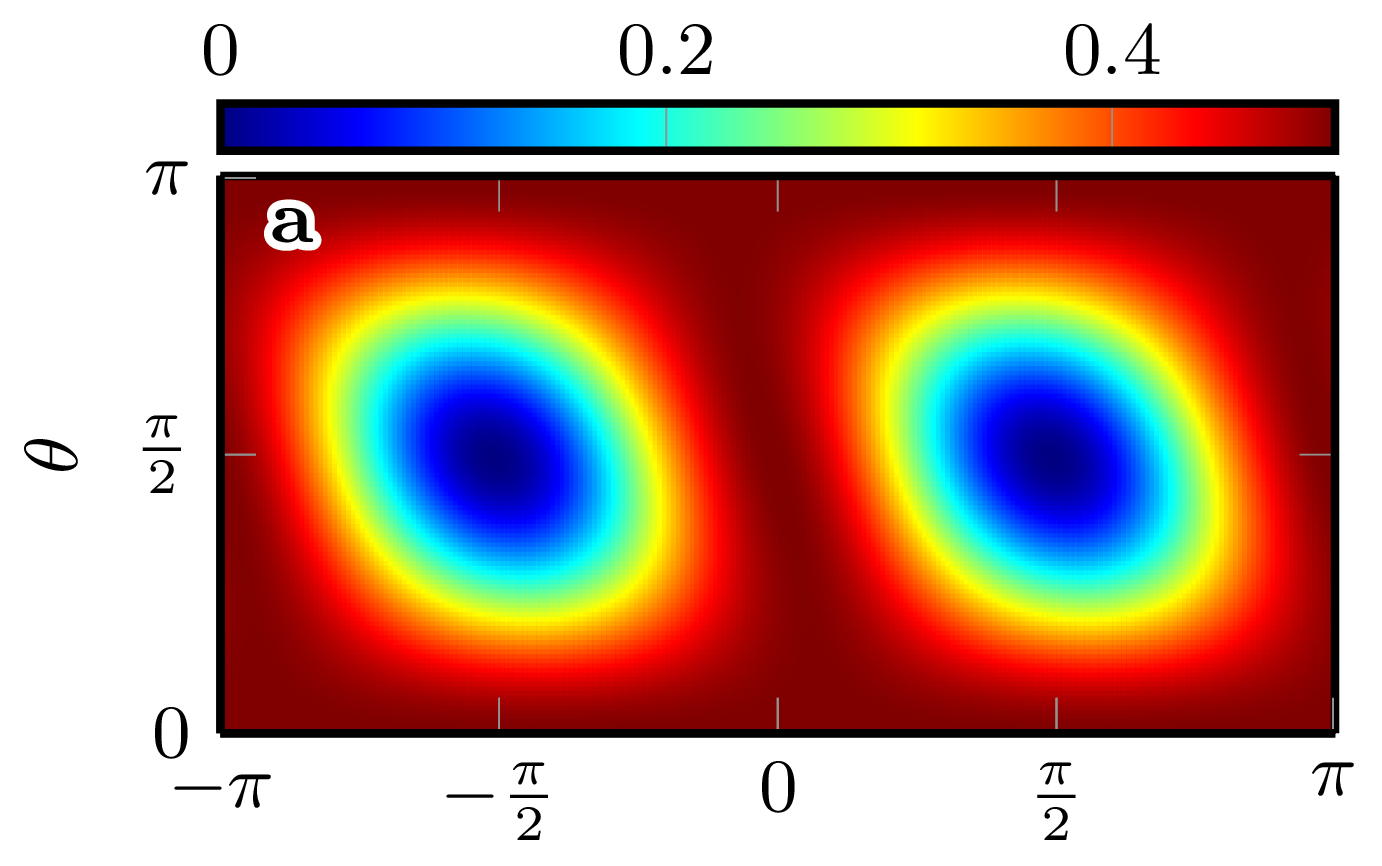}
    &
      \includegraphics[scale=\scaleval]{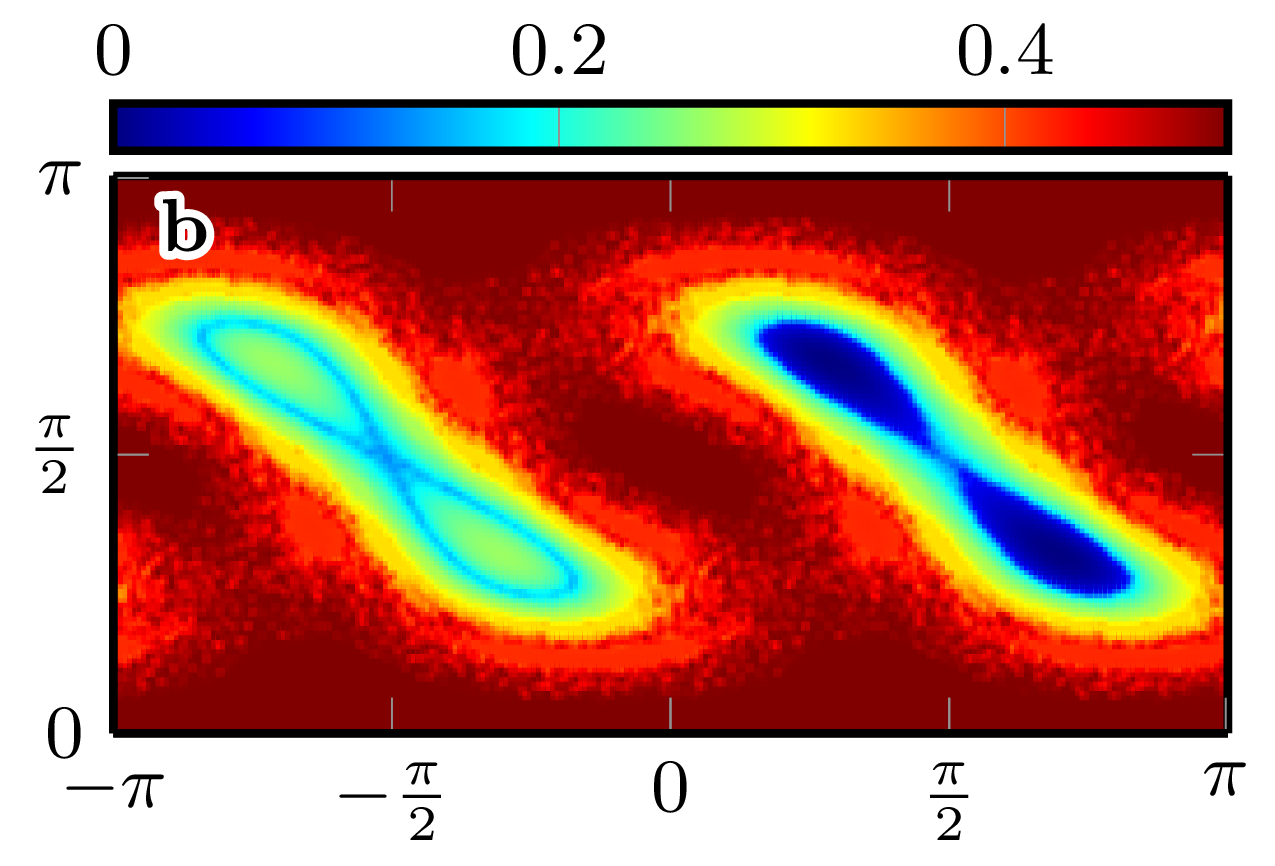}
    &
      \includegraphics[scale=\scaleval]{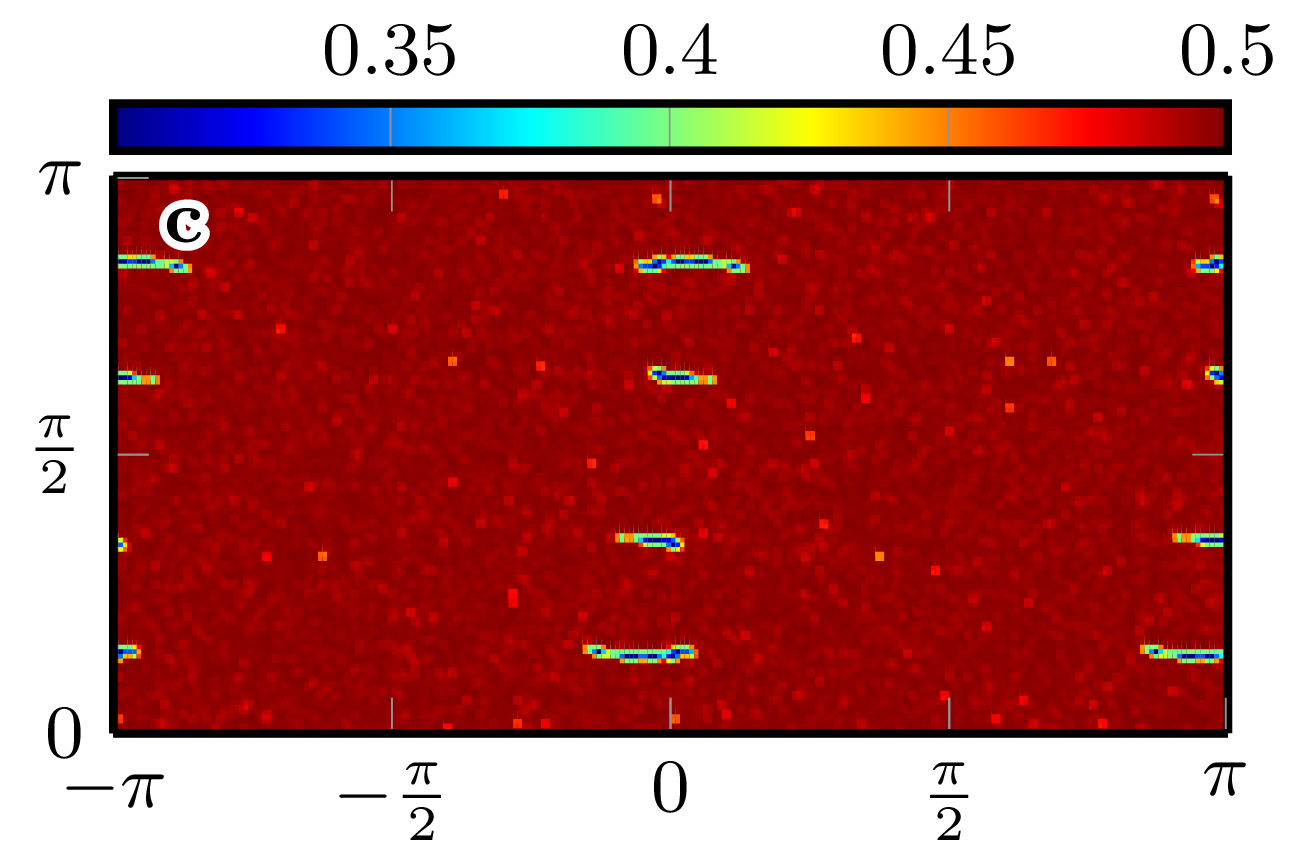}
    \\
    \includegraphics[scale=\scaleval]{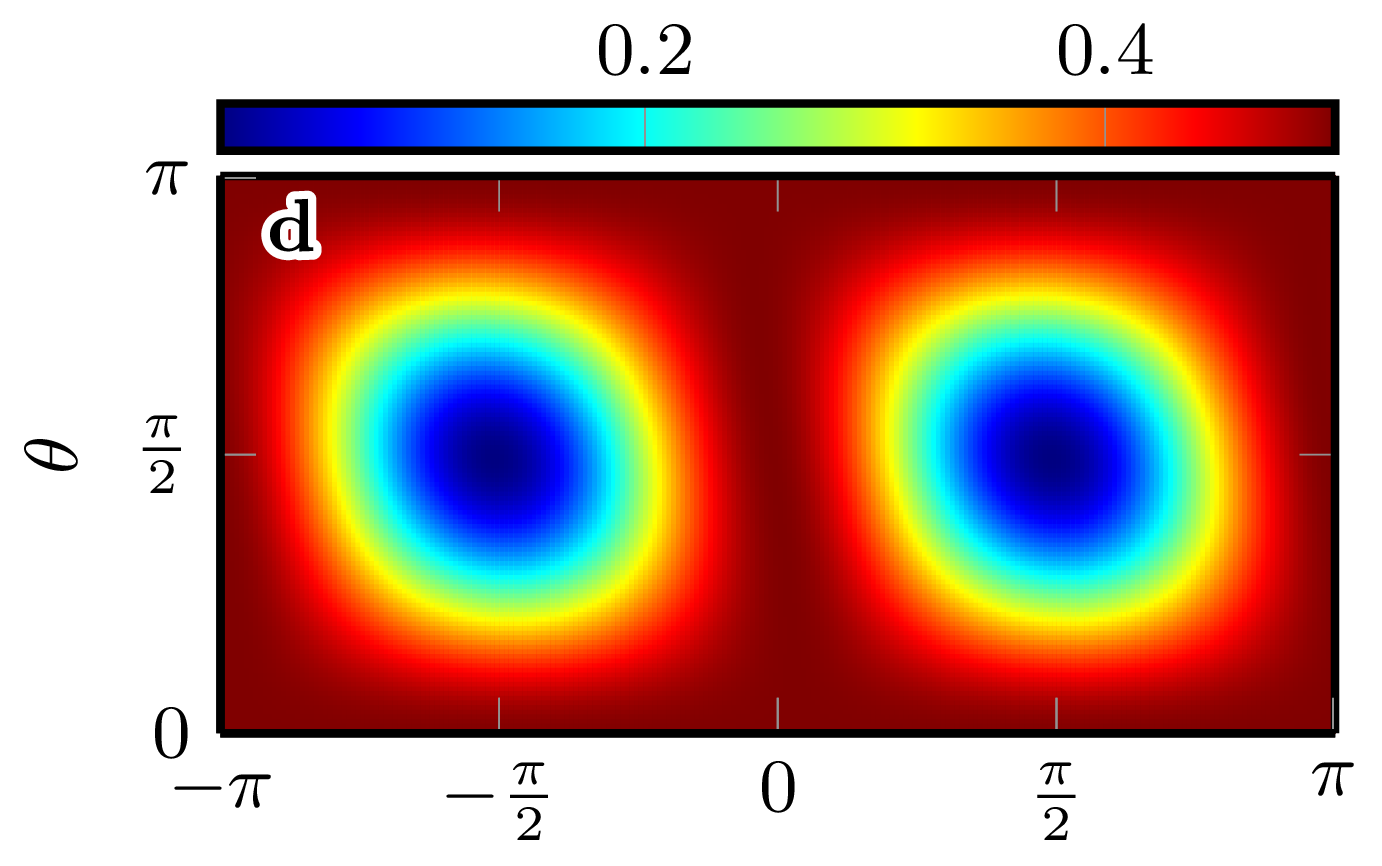}
    &
      \includegraphics[scale=\scaleval]{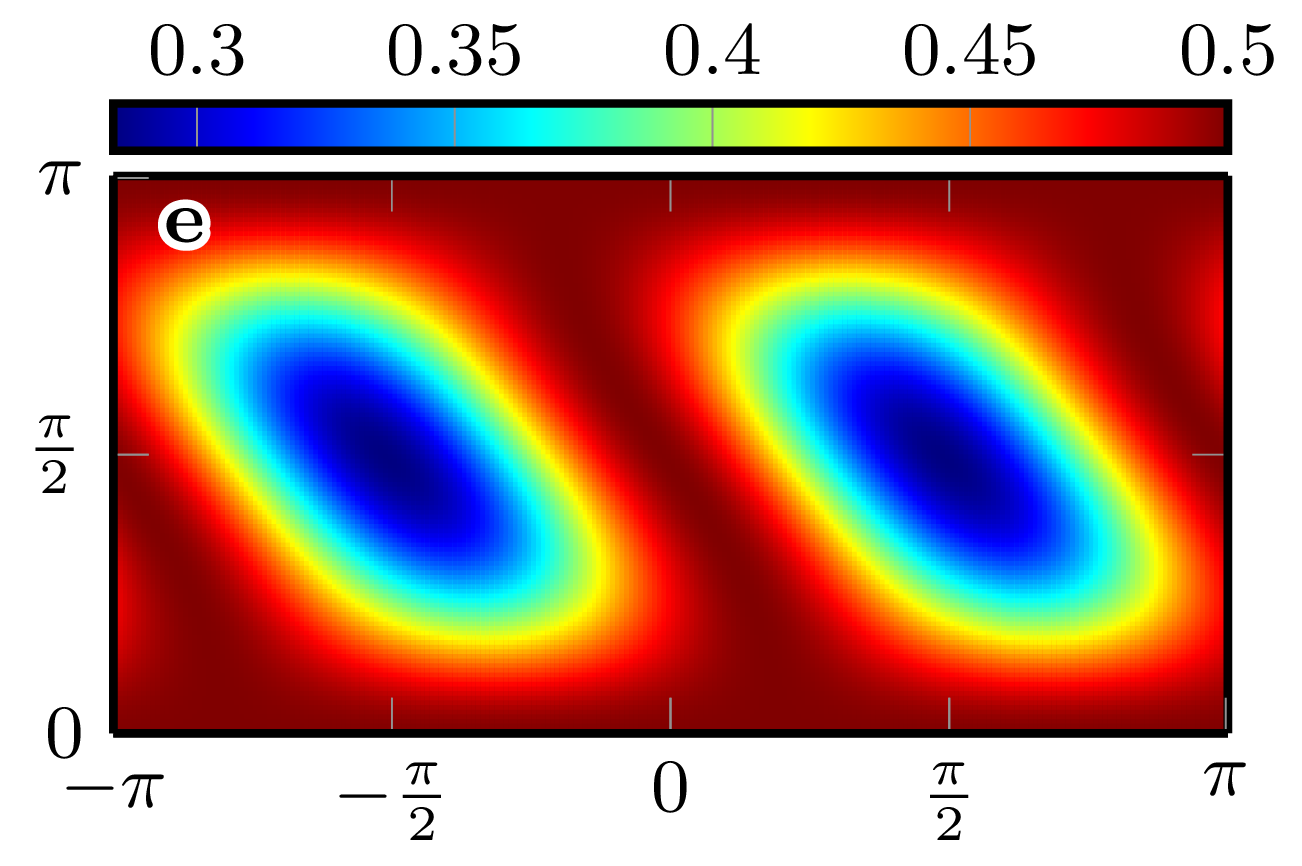}
    &
      \includegraphics[scale=\scaleval]{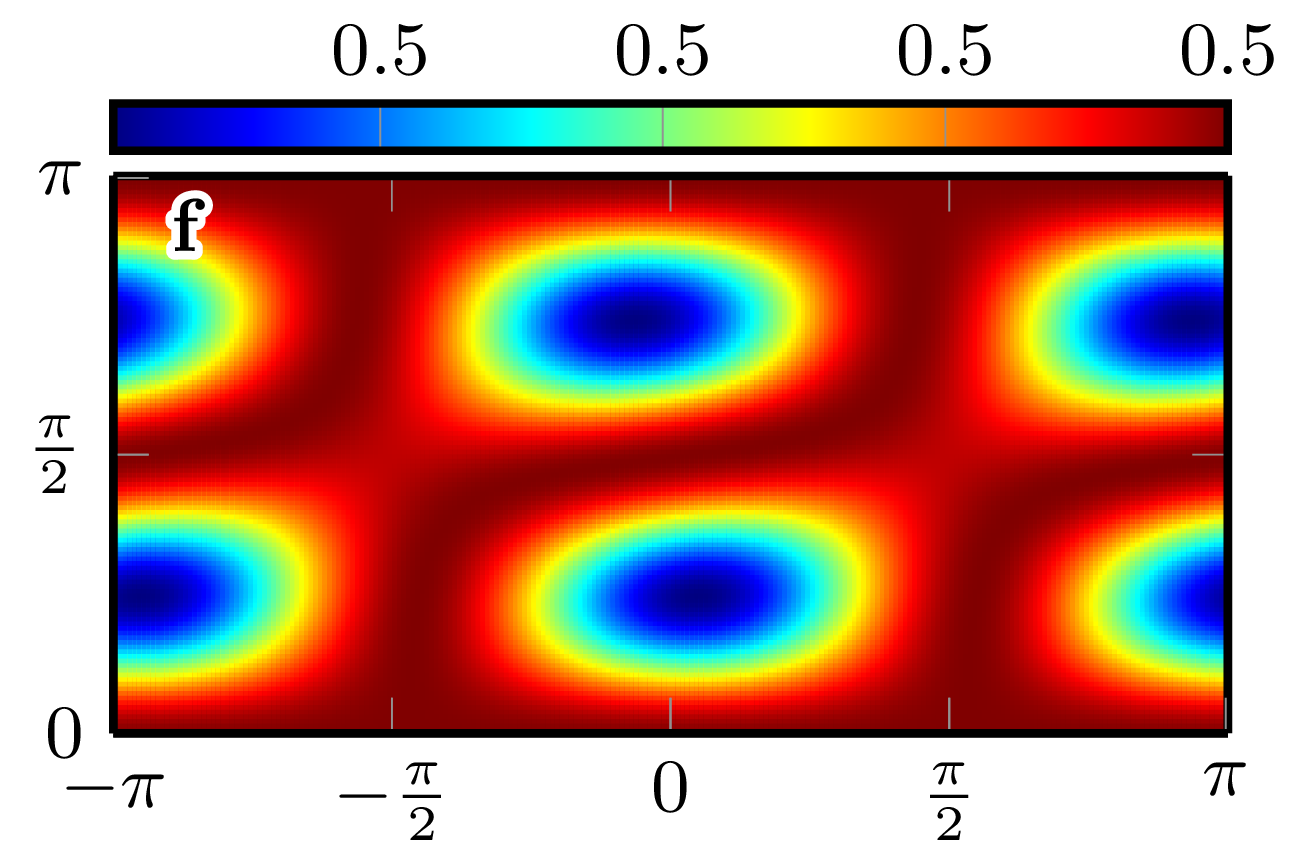}
    \\
    \includegraphics[scale=\scaleval]{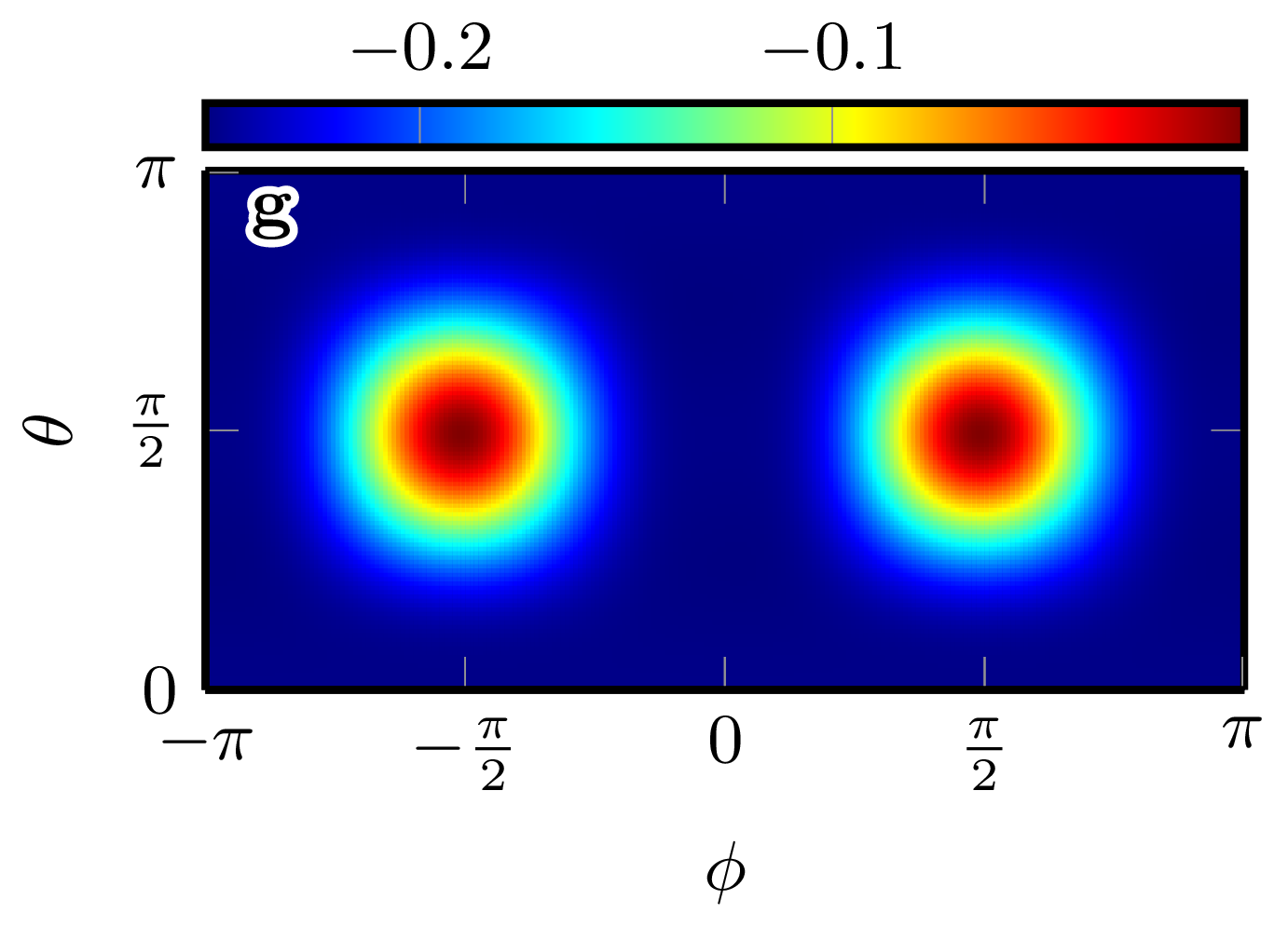}
    &
      \includegraphics[scale=\scaleval]{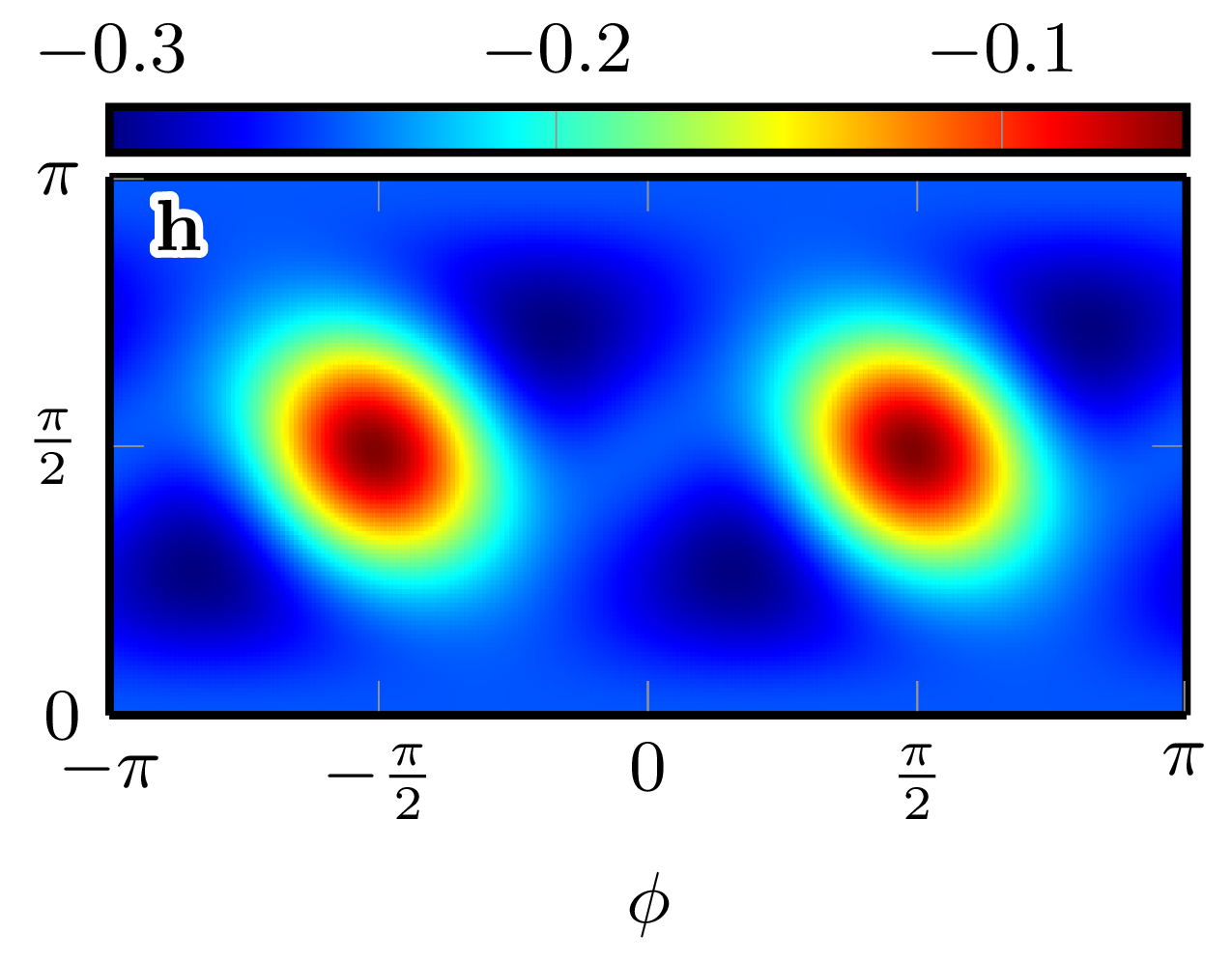}
    &
      \includegraphics[scale=\scaleval]{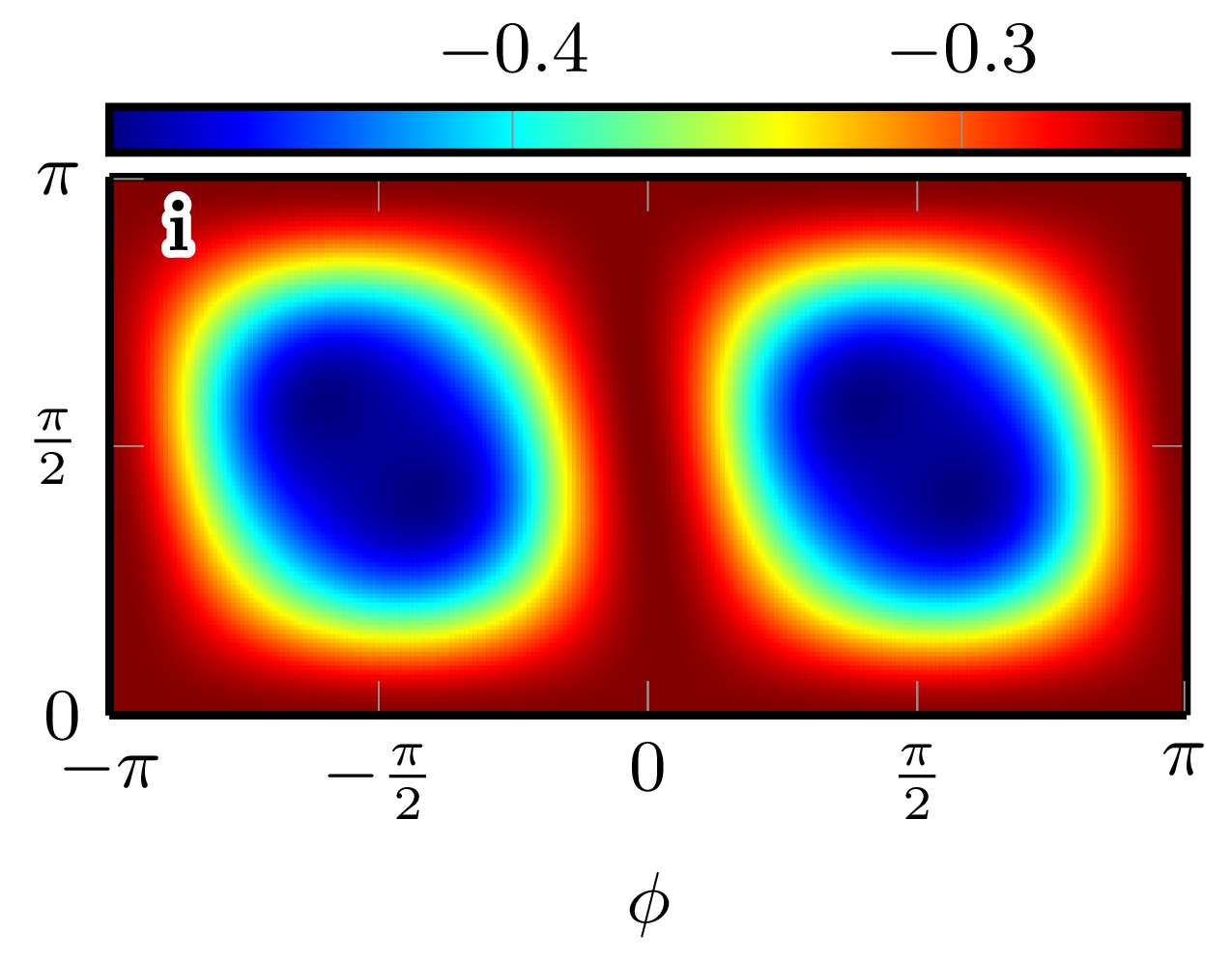}
  \end{tabular}
\caption{(a) Ignorance measure $I_C$ as a function of initial
condition $(\theta,\phi)$ for $\kappa=0.5$. 
(b) Same but for $\kappa=2.5$. (c) Same but for $\kappa=2\pi+0.5$. 
(d) Quantum ignorance measure $I_Q$ as a function of initial 
condition for $\kappa=0.5$. (e) Same but for $\kappa=2.5$. (f) Same but
for $\kappa=2\pi+0.5$.
(g) Quantum remainder term $R_Q$ as a function of initial 
condition for $\kappa=0.5$. (h) Same but for $\kappa=2.5$. (i) Same but
for $\kappa=2\pi+0.5$.
Note that although $S_Q$ (Fig.~\ref{fig:bigfig})
    has $\kappa$-period of $2\pi$, $I_Q$ and $R_Q$ both have period
    $4\pi$ and so do not have the same behavior at $\kappa=0.5$ and
    $\kappa=0.5+\pi$. Note the different scales for the different
    plots.}
  \label{fig:ignorance}
\end{figure*}

Since the similarities between the maps of initial-condition dependence 
of the entanglement and the classical Poincar\'e map cannot be explained 
by a direct connection between chaos and entanglement, we need to 
understand what other features of the dynamics renders the two so similar. 
In the following we argue that it is broad geometric features of the 
dynamics which provides this connection.

We start by considering that Eq.~\ref{E:entropy} can be rewritten as
\begin{align}
  S_Q &= I_Q+R_Q\\
I_Q &= \frac{1}{2}- \frac{1}{2j^2} \sum_i 
        \overline{\langle\psi(t)|\hat J_i|\psi(t)\rangle}^2\nonumber\\
  &= \frac{1}{2}- \frac{1}{2j^2} \sum_{k,q}E(k,k,q,q),\\
  R_Q &= -\frac{1}{2j^2} \sum_{k,l,p,q}\sum_i{C_k^*C_lC_p^*C_q}\langle \xi_k|\hat J_i|\xi_l\rangle
        \langle \xi_p|\hat J_i|\xi_q\rangle\nonumber\\
  &=-\frac{1}{2}\sum_{k,l,p,q}E(k,l,p,q)\nonumber\\
 &\text{ for } \{k,l,p,q:
\xi_{k}^*\xi_{l}\xi_{p}^*\xi_{q}=1, k\neq l \text{ or } p\neq q\}
\end{align}
with the overbar indicating time-averaging over an entire trajectory.
$I_Q$ is a global `ignorance' about the angular momentum for the
trajectory-averaged distribution associated with a given initial
condition.  It is essentially a measure of the delocalization across
the entire orbit, as the classical version below makes clear.  The
other `off-diaognal' remainder term is what we call $R_Q$, and has no
possible classical equivalent. These two measures are plotted in
Fig.~\ref{fig:ignorance}{\bf d}--{\bf f} and Fig.~\ref{fig:ignorance}
{\bf g}--{\bf i}, respectively. We conjecture that if a correlation
exists between $S_Q$ and a classical quantity it should be with
classical limit of $I_Q$. This classical quantity can be written as
\begin{equation}
  \label{eq:Ic}
  I_C=\frac{1}{2} - \frac{1}{2}(\overline{x(n)}^2+ \overline{y(n)}^2+\overline{z(n)}^2),
\end{equation}
which is plotted in Fig.~\ref{fig:ignorance}{\bf a}--{\bf c}. Using
this measure to compare with quantum dynamics incorporates the core
idea of Ref.~\cite{matzkin} that a classical orbit's delocalization is
relevant to the average entanglement of the corresponding quantum
state. However, this does not evoke a correspondence principle
assumption about the quantum and classical dynamics agreeing for any
length of time.  It also has the advantage of not needing the use of
classical distributions, from which it is computationally exceedingly
difficult to get converged results. Our conjecture relies on global
averages correlating even when local in time behaviors are
different. In fact, the plots of $I_C$ show a remarkable similarity to
the plots for $S_Q$ that persists for $\kappa=0.5,2.5$ at least until
the disconnect due to the quantum $\kappa$ periodicity seen at
$\kappa=0.5+2\pi$.


Before proceeding further in exploring this possibly useful approach, 
we need to strengthen the claim that the visual resemblance between 
the plots of $I_C$ and $S_Q$ gives us more insight than the visual 
resemblance of the $S_Q$ plots with the $\lambda$ plots. To do so, 
we need to quantify a distance measure between the various figures.

\subsection{Comparison of correlations}
\label{sec:comp-corr}

\begin{figure}
  \includegraphics[scale=\scaleval]{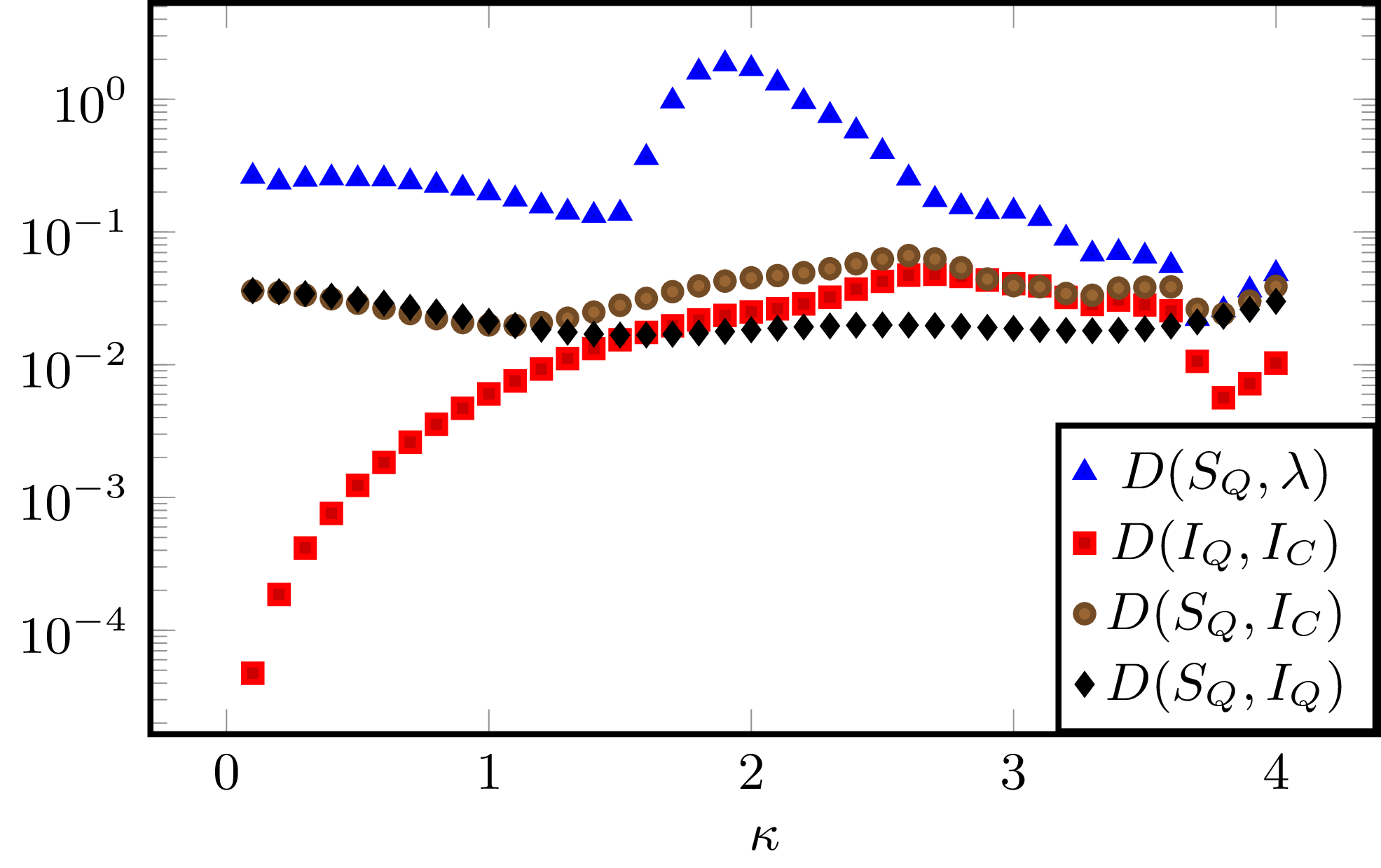}
  \caption{\label{F:distance}Correlation distances $D$ between
    entanglement and other measures. Values are only shown up to
    $\kappa=4$, which is where the break due to quantum periodicity
    occurs for two qubits.}
\end{figure}

To make concrete the visual similarities and differences of the many 
plots in Fig.~\ref{fig:bigfig}, we quantify the correlation 
between any two quantities $f(\theta,\phi)$ and $g(\theta,\phi)$ as
\begin{equation}
  D(f,g)=\ln[\frac{\Tr(f)\Tr(g)}{\Tr(fg)}],\label{eq:distance}
\end{equation}
a generalized Kullback-Liebler distance~\cite{prl2003}. Here $\Tr$
denotes the trace or double integral over the variables
$(\theta,\phi)$. A small distance $D$ implies good correlation, and
vice versa.  The behavior of this distance is shown in
Fig.~\ref{F:distance} for various quantities; note that we are
plotting $D$ on a logarithmic scale. This plot confirms that indeed
$D(S_Q,\lambda)$ is large---that is, the entanglement and chaos are
uncorrelated or weakly correlated. On the other hand, $S_Q$, $I_Q$,
and $I_C$ all show good correlation with one another, supporting our
conjecture that if there is a connection between $S_Q$ and a classical
measure, it should be our ignorance or delocalization measure $I_C$.
The extremely good correlation between $I_Q$ and $I_C$ for $\kappa<1$
is due to the fact that the limits $j\to\infty$ and $\kappa\to0$ are
related, as can be seen from the Hamiltonian
(Eq.~\ref{eq:hamiltonian}).

A source of disagreement (for larger $\kappa$ values) between the 
classical and quantum measures is due to the $\kappa$-periodicity 
of the quantum dynamics as noted in Sec.~\ref{sec:entangl-dynam-2}. 
Thus the quantum and classical geometries part company as the 
quantum system cycles through different behaviors in contrast to the 
increasing classical chaos. In general, this implies a distinct break 
in the quantum-classical connection after a specific 
$\kappa^{\text{max}}(2j)$ depending on the number of qubits $2j$. 
Thus, when evaluating the correlation of various quantities
we focus on $\kappa<\kappa^{\text{max}}$.

We also see that although the entirely quantum terms comprising $R_Q$
can be quite large compared to $I_Q$, this does not seem to
significantly affect the correlation seen between $S_Q$ and $I_C$.  A
clue to this may perhaps be found in the fact that although $R_Q$ has
no classical analog, its limiting behavior is determined by $S_Q$
and $I_Q$. As $j \to \infty$ we must have $S_Q\to0$ and $I_Q\to
I_C$. Since $S_Q=I_Q+R_Q$, it therefore must be true that $R_Q\to-I_C$
in the classical limit. Thus, arguably any global reason for the
correlation between $S_Q$ and $I_C$ could also apply to $R_Q$,
although it is difficult to extend this further given that there is no
classical analog for $R_Q$. Consequently, we focus below on $S_C$ and
$I_C$.

\section{Shared symmetry}
\label{sec:shared-symmetry}

\begin{figure}
  \centering
  \includegraphics[scale=\scaleval]{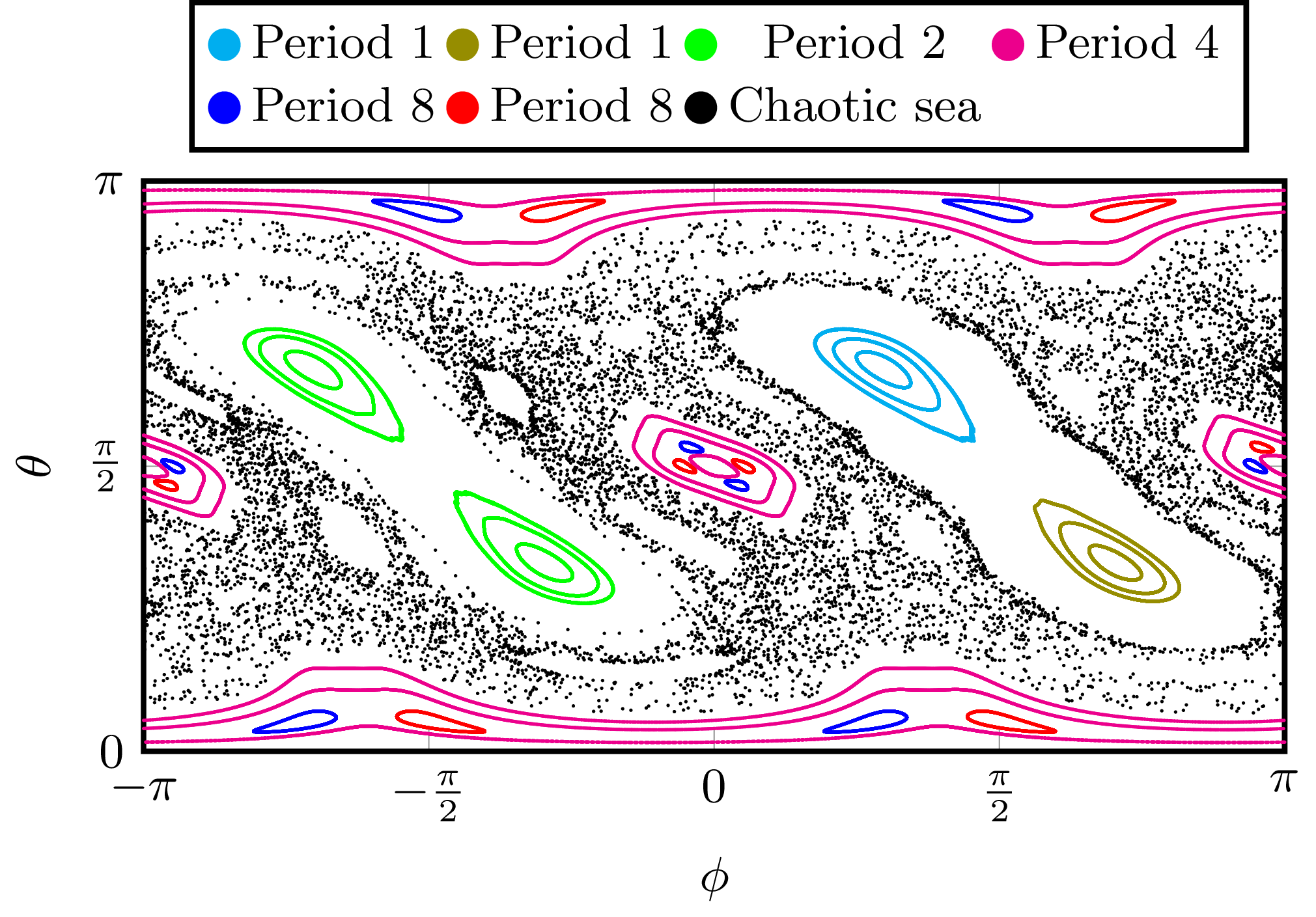}
  \caption{\label{F:periods} Periodic orbits and surrounding stable islands 
    for the classical system at $\kappa=2.5$.}
\end{figure}

\begin{figure}
  \centering
  \begin{tabular}{c}
  \includegraphics[scale=\scaleval]{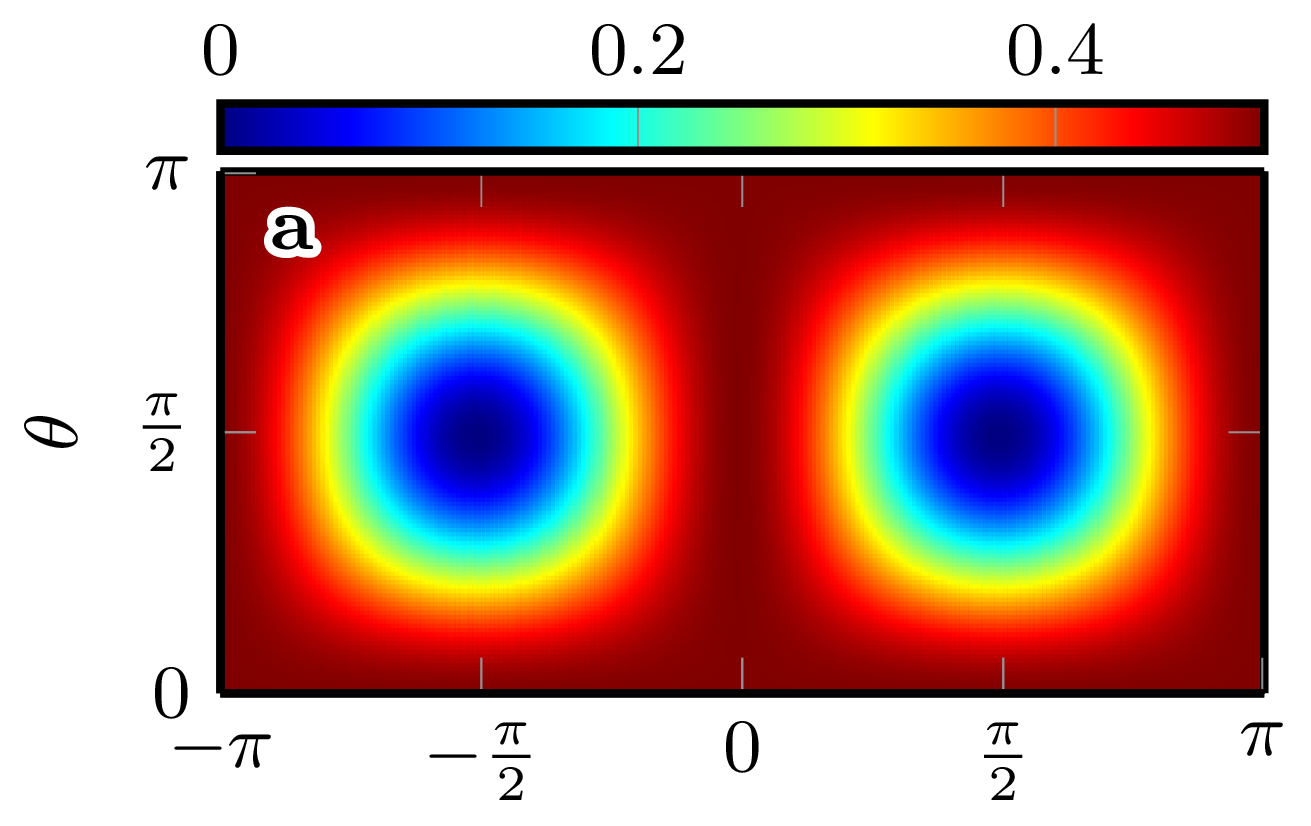}\\
  \includegraphics[scale=\scaleval]{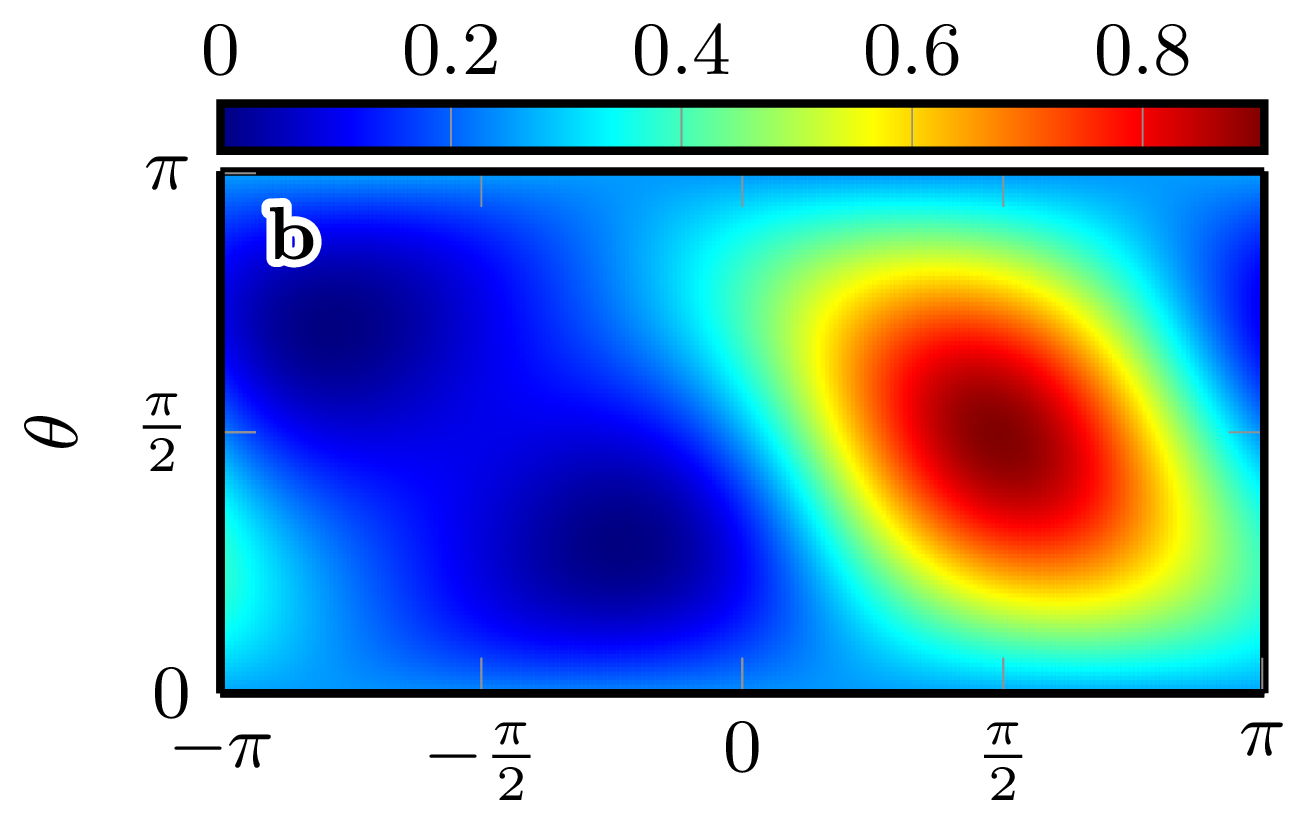}\\
  \includegraphics[scale=\scaleval]{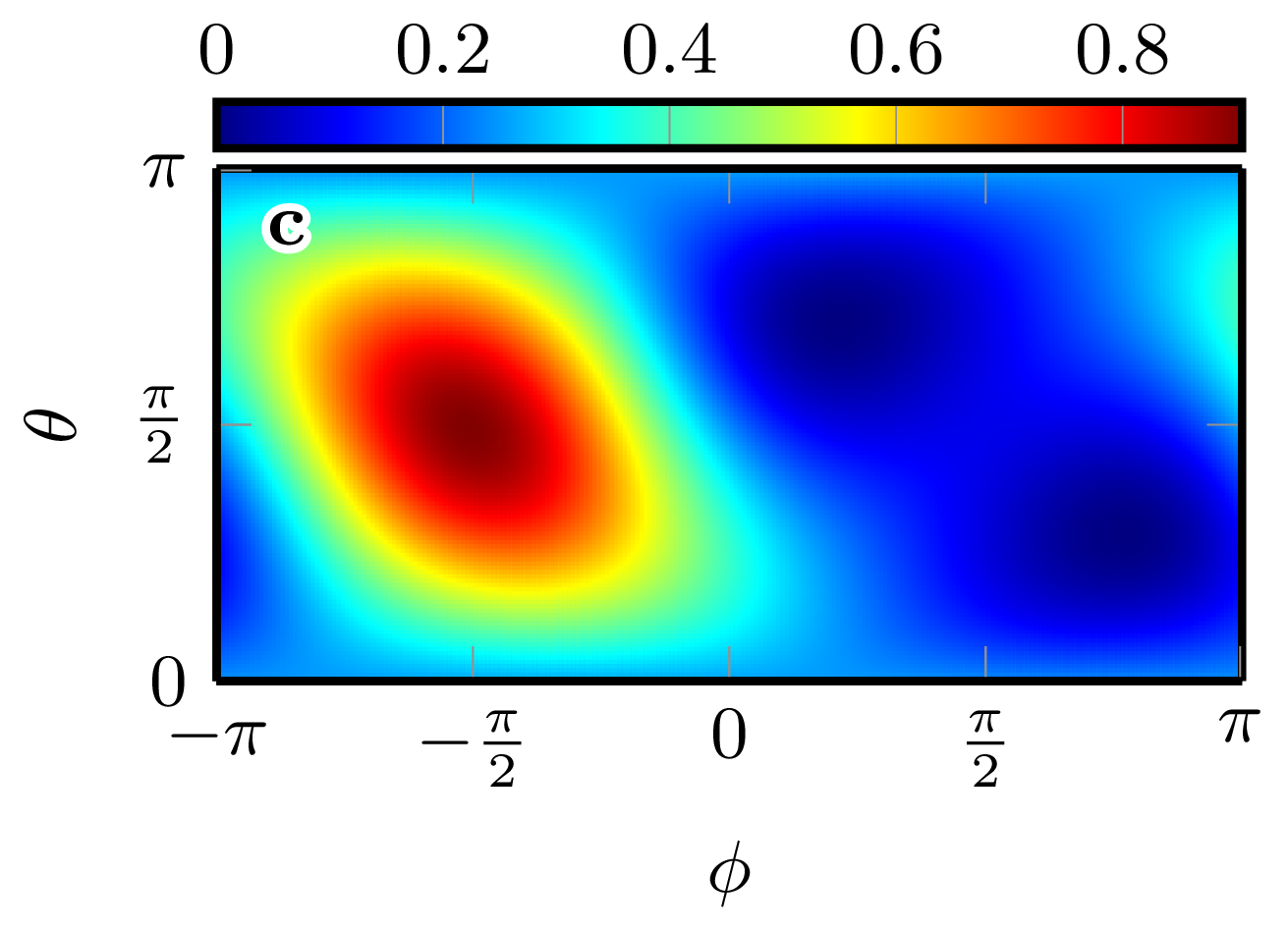}
  \end{tabular}
  \caption{The Husimi phase-space representation for the three eigenstates 
    of the Floquet operator $U$.}
  \label{F:eigenst}
\end{figure}

\begin{figure}
  \begin{tabular}{c c}
    \includegraphics[scale=\scaleval]{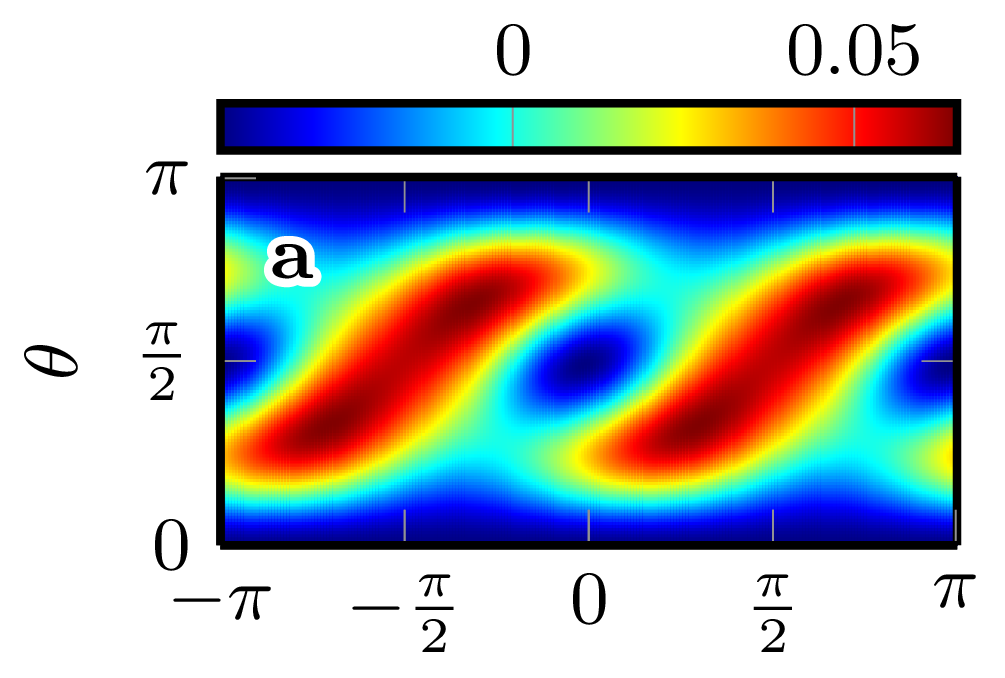}
    &\\
    \includegraphics[scale=\scaleval]{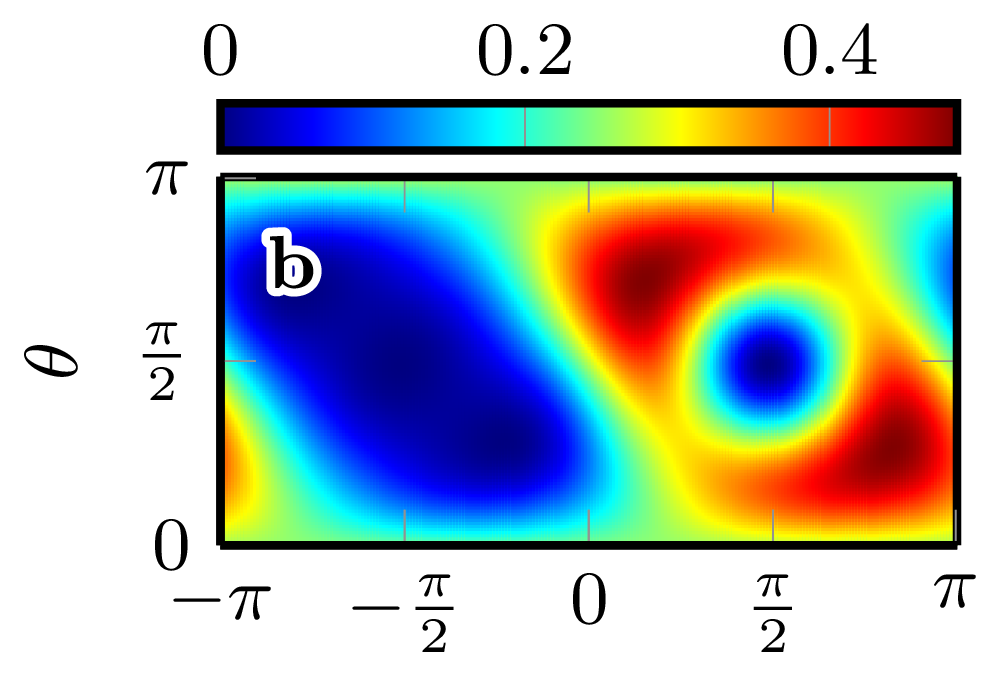}
    &
    \includegraphics[scale=\scaleval]{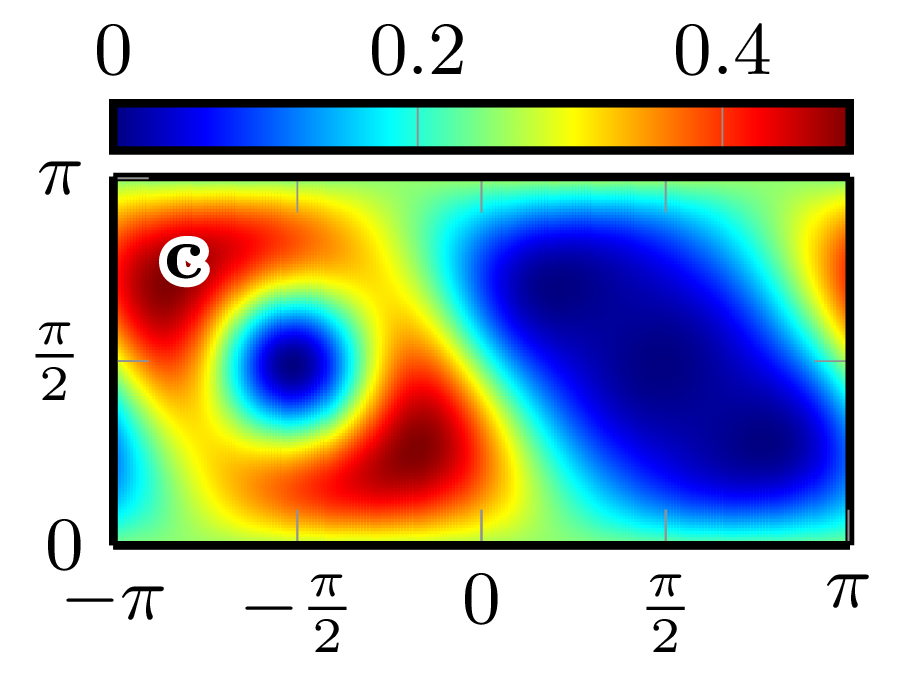}
    \\
    \includegraphics[scale=\scaleval]{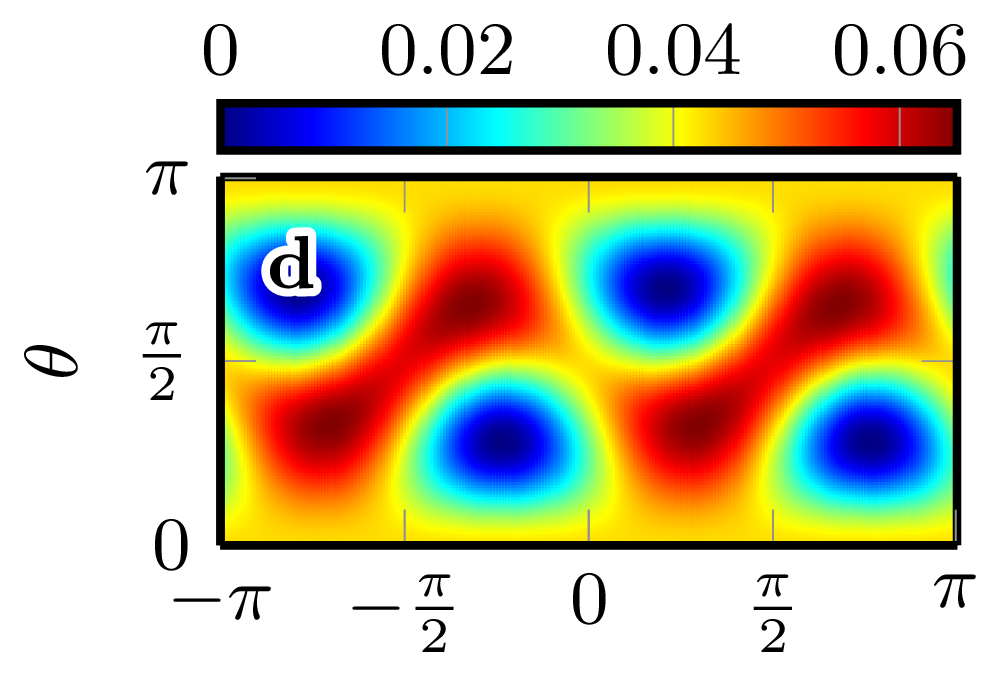}
    &
    \includegraphics[scale=\scaleval]{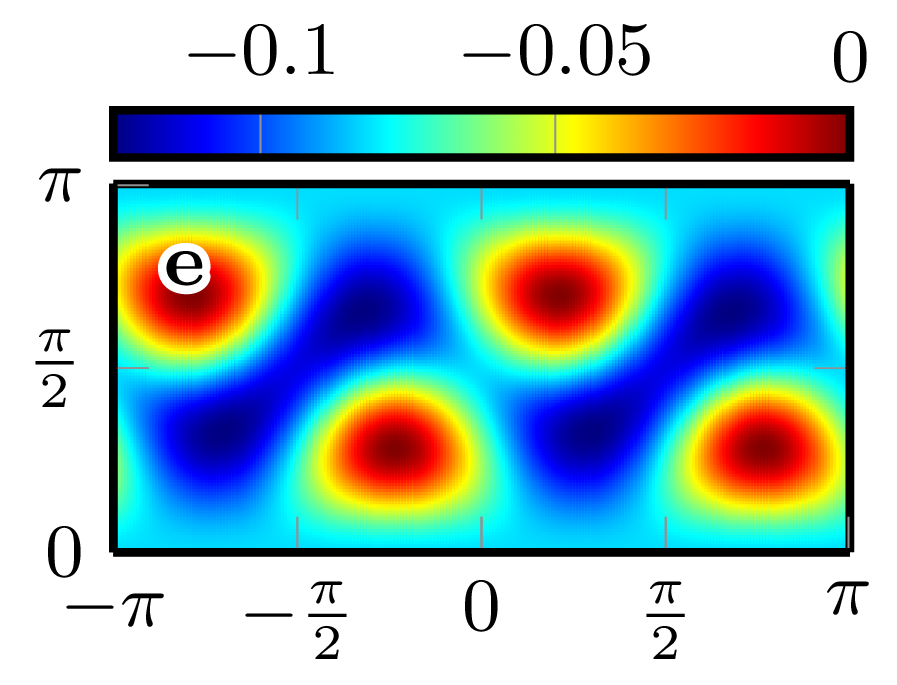}
    \\
    \includegraphics[scale=\scaleval]{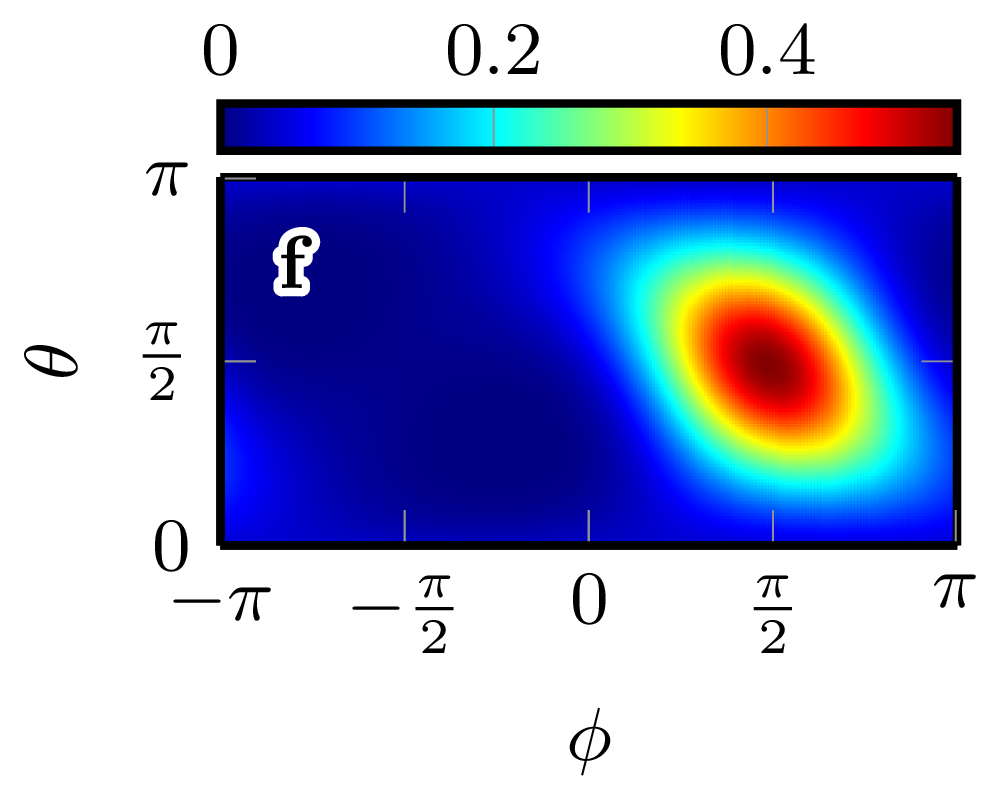}
    &
    \includegraphics[scale=\scaleval]{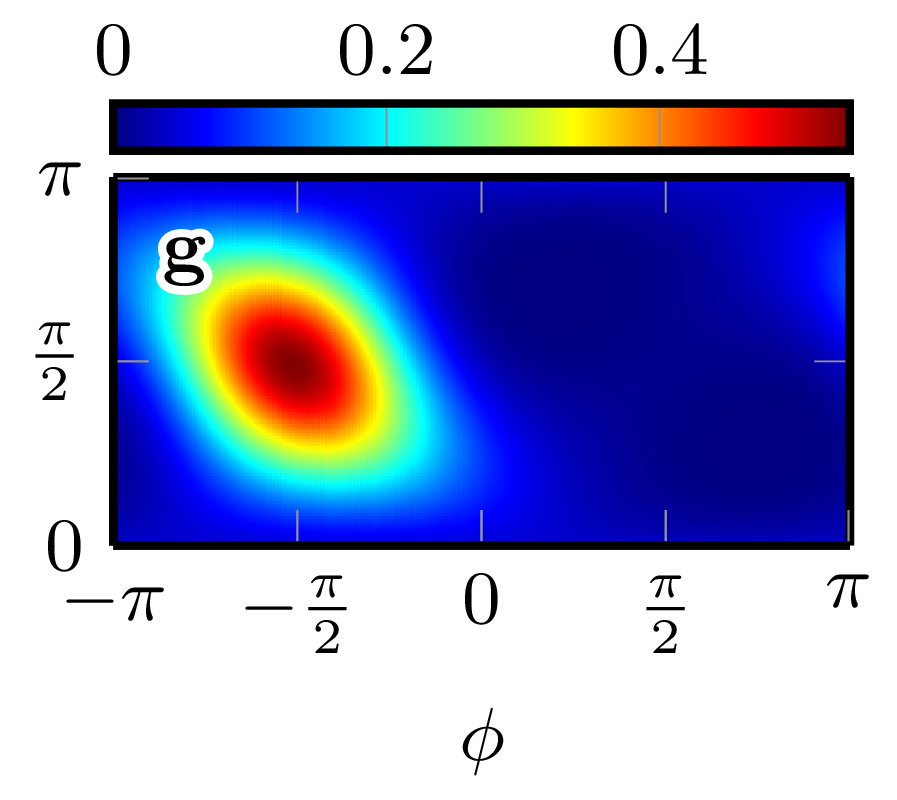}
  \end{tabular}
  \caption{\label{F:eigencomp} Compact representation of the non-zero
terms from the $81$ different $E(k,l,p,q)$: we add the $E(k,l,p,q)$ that 
are identical or conjugates of one another. These are 
plotted as a function of $\theta$, $\phi$, at $\kappa=2.5$. 
(a) The sum of $E(2,3,2,3) + E(3,2,3,2)$ (which are conjugates of each
other, thus yielding a real result);
(b) the sum $E(1,2,2,1)+ E(2,1,1,2)$ (which are identical to each other);
(c) the sum $E(1,3,3,1)+ E(3,1,1,3)$ (which are identical to each other);
(d) the sum $E(2,3,3,2)+ E(3,2,2,3)$ (which are identical to each other);
(e) the sum $E(2,2,3,3)+ E(3,3,2,2)$ (which are identical to each other);
(f) $E(2,2,2,2)$
(g) $E(3,3,3,3)$.
Thus panels (b)-(e) sum to $R_Q$, and (a), (f), and (g) sum to $I_Q$. 
Note the different scales on each plot.}
\end{figure}
Some insights about these similarities that we have observed obtain from
considering the symmetries associated with both the classical and
quantum dynamics. Specifically, there are four classical
symmetries~\cite{haake} associated with the kicked top dynamics. These
arise from the fact that the classical map $F$ is invariant under two
non-standard time reversals and rotation by $\pi$ about the $y$ axis,
and $F^2$ is invariant under rotations by $\pi$ about the
$x$ axis. Exact analogues of all four symmetries exist in the quantum
map $\hat U$ (see Ref.~\cite{haake} for an extended discussion of
these symmetries).

The relevant aspect of these symmetries for our consideration here is
to consider how different classical orbits organize.  In
Fig.~\ref{F:periods} we plot the classical periodic orbits of
different period, along with the associated islands of stability. The
locations of these periodic orbits are determined by the classical
symmetries. The actual locations of the periodic orbits change with
$\kappa$, while the size of the stability islands around the periodic
orbits also shrinks with increasing $\kappa$.  Further, we expect that
the Floquet eigenstates of $\hat U$ naturally carry the same dynamical
symmetries as the quantum map. While this is true
(Fig.~\ref{F:eigenst}), what is striking about these states is that
they seem far too large to resolve the smaller island structures seen
in the classical phase space and the quantum entanglement figures.

We can unfold this apparent paradox by focusing attention on the
actual operator averages we need to compute. As we see in
Fig.~\ref{F:eigencomp}, the dynamical symmetries of the quantum map
(which are shared with the classical map) are also reflected in the
plots of the various $E(k,l,p,q)$ that sum to $I_Q$ (diagonal terms) and
$R_Q$ ('off-diagonal' terms).  That is, the ability to resolve
smaller-scale structures for the $E(k,l,p,q)$ is more critical than the
more spread-out shape of the eigenfunctions themselves.
Comparing Fig.~\ref{F:eigencomp} of the various $E(k,l,p,q)$ 
with Fig.~\ref{F:periods},
and further comparing these plots with all the plots in
Fig.~\ref{fig:bigfig} also makes clear that (a) both classical and quantum
dynamics show signatures of the same symmetries and (b) quantities
such as $E(k,l,p,q)$ and consequently $S_Q$ as well as $I_C$ reflect 
these phase-space symmetries. This leads to our argument that the 
long-observed correlation between measures of the classical and quantum 
systems arises from both kinds of phase space being organized around the
symmetries of the dynamical system rather than any particular
dynamical property such as the classical trajectories' degree of
chaos. Since classical stability islands and chaotic `seas' also
organize around phase-space symmetries, this explains how a seeming
association between chaos and entanglement can appear.

Focusing on the symmetries also allows us to separate the behaviors of
those classical regular orbits that correspond to the highest quantum
entanglement from those which correspond the lowest
entanglement---they are indeed orbits of very different symmetries.
The organization of $I_C$ around symmetry points is clear: points that
are invariant under symmetries have $I_C=0$, and those whose
relationship to the symmetries causes them to move between a few small
but distant areas on the sphere have maximal $I_C$. To relate the
quantum entanglement to the symmetries, we can adapt the argument
of Ref.~\cite{matzkin} to relate each of these to the spread of the quantum
state. Since total angular momentum is conserved,
Eq.~\ref{eq:entropy1} can be rewritten in terms of the variances
$\sigma_i=\sqrt{\langle J_i^2\rangle - \langle J_i\rangle^2}$ in order
to show that a state with higher spread is more entangled. Then, the
initial states whose relationship to the symmetries is such that they
get highly spread out over the sphere have a much higher average
entanglement; intuitively, these are exactly the same initial
conditions that classically end up with high values of $I_C$.

One effect that is not visible in the classical Poincar\'e map jumps
out if the specific time-period is noted for different periodic orbits
(Fig.~\ref{F:periods}). For example, the largest islands are
associated with a period-2 orbit for negative $\phi$. This, however,
looks very similar to the two period-1 orbits for positive
$\phi$. This explains the corresponding symmetry breaking in $I_C$
(Fig.~\ref{fig:ignorance}{\bf a}--{\bf c}). This symmetry breaking is not
observed in any of the other plots of Fig.~\ref{fig:bigfig}. However,
this symmetry breaking was observed in plots of $S_Q$ in a 3-qubit
experiment~\cite{pedram}. We may legitimately conjecture that this
emerges in the experimental entanglement measures due to the dynamical
difference of the symmetry breaking being enhanced by experimental
noise or decoherence, although modeling that is beyond the scope of
this paper.

\section{Discussion}
There are several points worth noting about the semiclassical and 
high $\kappa$ limits of this analysis, although both remain out of the
scope of this paper.

We have remained focused on the observation that the most extremely
quantum system shows shapes in the map of $S_Q$ that resembles shapes
in the classical phase space, albeit appearing as either correlation
or anti-correlation between classical regularity and high or low
quantum entanglement.  Arguments previously advanced in the
literature, particularly in the semiclassical limit, cannot
apply. That is, {\em a priori} there cannot be a simple link between
two-qubit dynamics and classical dynamics, even while the connection
evidenced by the resolution of the $S_Q$ plots invites an explanation.
It seems clear, however, that the arguments we advance about symmetry
and the $I_Q\to I_C$ connection should hold in the semiclassical
case. This suggests that consideration of eigenstates which have 
increasingly sharper support in phase space at higher $j$ values will 
not alter the overall relationship between quantum and classical 
dynamics.  

Wang {\it et al}. \cite{wang} have argued that in the semiclassical regime,
entanglement and chaos show similar dependence on increasing $\kappa$
once the system has become fully chaotic ($\kappa>3.5$).  However, due
to the small $\kappa$ periodicity of the 2-qubit quantum system, we do
not expect any connection at all between the classical and quantum
systems after $\kappa\approx3$. Thus, we cannot examine the fully
chaotic classical system in the context of our system.

\begin{figure}
  \includegraphics[scale=\scaleval]{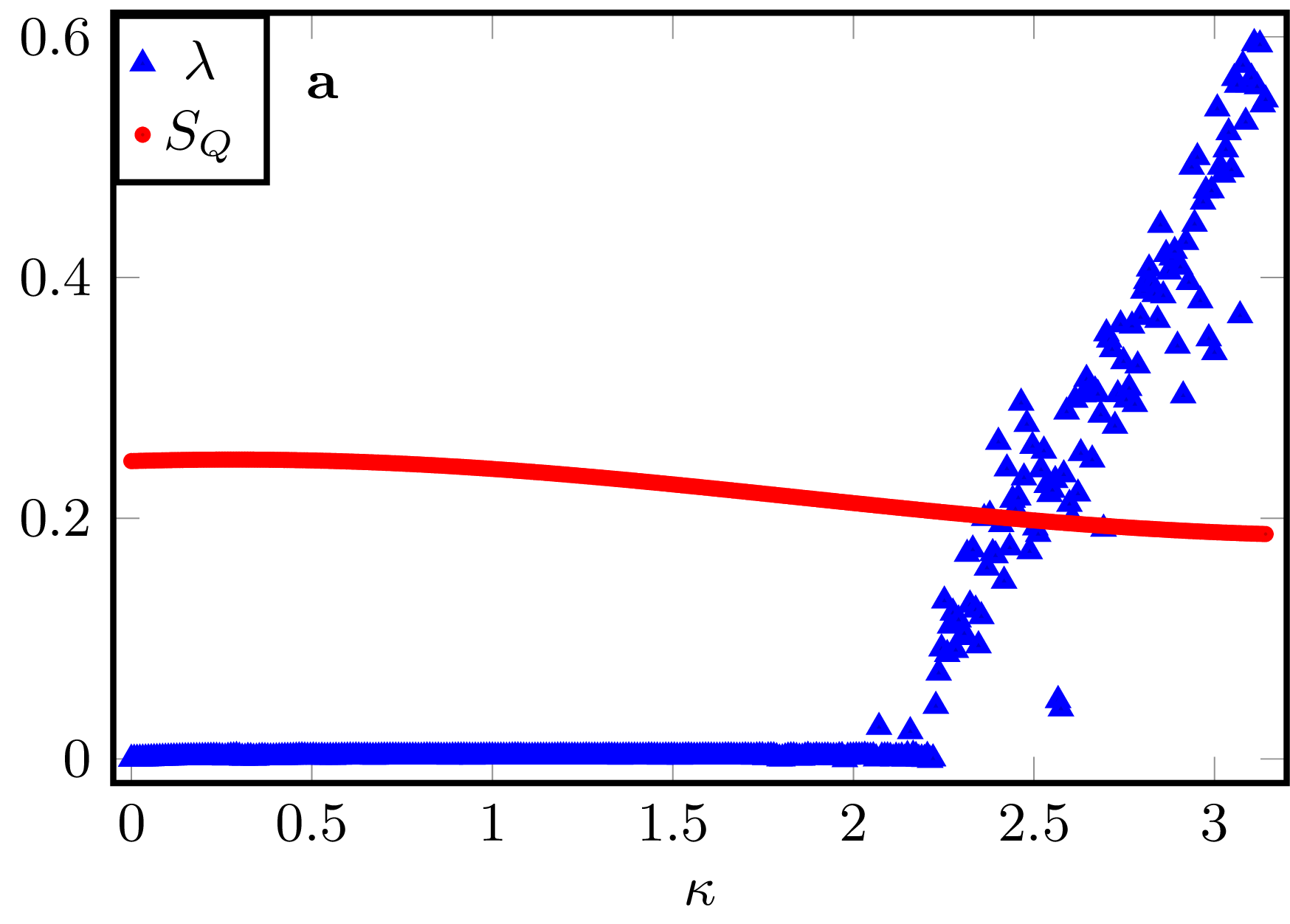}
  \includegraphics[scale=\scaleval]{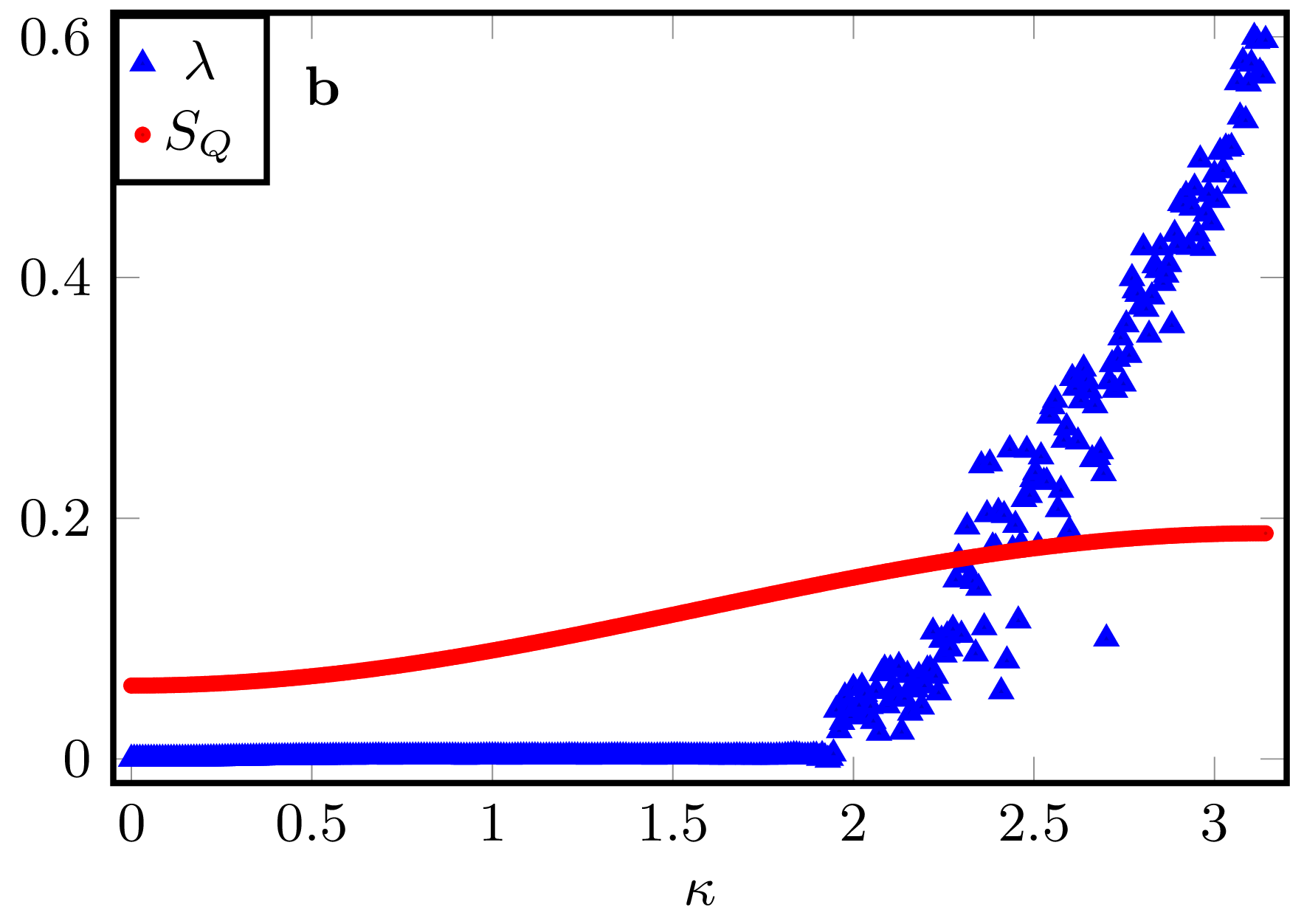}
  \caption{(a) A plot of time-averaged entanglement $S_Q$ and Lyapunov
    exponent $\lambda$ as a function of $\kappa$ with initial conditions, 
$(\theta,\phi) =(2.35,-0.1)$. 
(b)Same but with $(\theta,\phi)= (2.35,-\pi/2)$. These two figures 
demonstrate that although we can pick an initial condition so
entanglement entropy and Lyapunov exponent appear correlated, 
this is not true in general.} 
  \label{fig:funcofkappa}
\end{figure}

Instead, we can explore the following. At higher $\kappa$ values, the
stability islands surrounding the periodic orbits become vanishingly
small and the phase space is dominated by chaotic orbits. All chaotic
orbits explore phase space in essentially similar ways and in the
absence of the periodic orbits that lead to either high or low
entanglement, there should be little variation in the initial
condition dependence of entanglement.  However, in considering {\em
  only} the chaotic region, we have checked to see if at a given
location and with varying $\kappa$ there is any similarity between how
$\lambda$ changes compared with how $S_Q$ changes. In
Fig.~\ref{fig:funcofkappa} we show examples of our findings that there
is essentially no correlation between the $\kappa$ dependence of
$\lambda$ and $S_Q$: all initial conditions show essentially the same
behavior for $\lambda$, and we can choose an initial condition to find
essentially any behavior we like for $S_Q$.  Some previous studies
\cite{wang,Miller} have investigated the time-dependence of the
entanglement in the semiclassical regime and found that quickly
entangling initial conditions correspond with classically chaotic
regions. Due to the very short period observed in the two-qubit case,
the rate of entanglement is not a meaningful quantity in this study,
and so we focus on the initial-condition dependence.

None of this would disagree with correspondence between quantum and
classical behavior for the fully chaotic system in
particular, as has been achieved using random-matrix theory~\cite{RMT}.

We additionally note that both the symmetry observations and
definitions of our ignorance measures are reliant on the spherical
geometry of the phase-space of the kicked top. It is unclear whether
there exist analogues for other geometries, and thus whether such a
connection between quantum entanglement and classical phase-space
maintains in other geometries. The restriction of this system to the
symmetric subspace (via angular momentum conservation) is also
essential our argument concerning the ignorance measures, since it
gives the equivalence between entanglement and delocalization of the
quantum state.

\section{Conclusion}
\label{sec:conclusion}
We have demonstrated through numerical and analytical calculations
consistent with recent experiments that entanglement, a quintessentially
quantum phenomenon, is associated with the geometry of the classical
phase space dynamics through the symmetries of the shared
Hamiltonian. There is also a connection between the time-averaged
entanglement of the quantum system and the ignorance (effectively 
delocalization) measures $I_Q$ and $I_C$. We have seen this connection 
for two-qubit systems (the most quantum regime possible). We also report that 
the entanglement dynamics are periodic or quasi-periodic in time depending
on the nonlinearity parameter $\kappa$, as well as being a periodic
function of $\kappa$. All of these results generalize to higher
numbers of qubits.

There are several interesting directions in which this connection
between entanglement and dynamical nonlinearity could be explored. 
The first is to understand better how the behavior changes as $j$
increases. While we do discuss in general terms how increasing
$j$ works, a detailed analysis explicitly linking the very high $j$
and the low $j$ systems would be illuminating (albeit ambitious),
particularly with regard to understanding the differences between
finite-time and infinite-time averages of various quantities.
A second approach is to understand the initial condition dependence of
out-of-time-ordered-correlators~\cite{swingle} and determine
whether this bridges some notions of quantum and classical information
loss due to dynamics. It would also be worthwhile to understand the
precise source of the differences in the quantum behavior of islands
associated with different periods; as pointed out above, this
classical difference remarkably seems visible in the three-qubit
experiment~\cite{pedram} but not in the quantum calculations for
either two-qubit or three-qubit systems. These should all help further
clarify the relationship between entanglement and classical dynamics.

\section*{Acknowledgements}
We are grateful for funding from the Howard Hughes Medical Institute
through Carleton College. We also acknowledge useful discussions with
Pedram Roushan, Charles Neill, as well as Poul Jessen and helpful feedback 
from an anonymous referee.


\begin{thebibliography}{99}
\bibitem{fn1}Although strictly this notation assumes nondegenerate 
eigenvalues, this is neither necessarily true or needed and we use 
it here only for convenience.
\bibitem{fn2}Plots from
  previous studies~\cite{wang} are consistent with this result, but
  their presentation via contour lines makes this particular
  relationship indistinguishable from strict correspondence between
  chaos and entanglement.
\end{thebibliography}

\end{document}